\documentclass[prb,twocolumn,preprintnumbers,amsmath,amssymb]{revtex4}
\usepackage{graphicx}
\usepackage{color}
\usepackage{multirow}
\usepackage[sort&compress]{natbib}
\newcommand{\bra}[1]{\ensuremath{\langle #1 |}}
\newcommand{\ket}[1]{\ensuremath{| #1 \rangle}}
\newcommand{\meV}{\ensuremath{\,\mbox{meV}}}

\newcommand{\mueV}{\ensuremath{\,\mu\mbox{eV}}}

\newcommand{\nm}{\ensuremath{\,\mbox{nm}}}

\newcommand{\mx}[1]{\ensuremath{\widetilde{#1}}}
\newcommand{\vc}[1]{\ensuremath{\vec{#1}}}

\newcommand{\Bmb}{\ensuremath{\mathcal{B}^{\mathrm{MB}}}}
\newcommand{\Hop}{\ensuremath{\mathcal{H}}}
\newcommand{\Hmb}{\ensuremath{\mathcal{H}}}
\newcommand{\trsigma}{\ensuremath{\sigma}}

\begin{document}

\title{A configuration interaction analysis of exchange in double quantum dots}
\author{Erik Nielsen and Richard P.~Muller}
\affiliation{Sandia National Laboratories, Albuquerque, New Mexico 87185 USA}
\date{\today}

\begin{abstract}
We describe in detail a full configuration interaction (CI) method designed to analyze systems of quantum dots.  This method is capable of exploring large regions of parameter space, like more approximate approaches such as Heitler London and Hund Mulliken, though it is not limited to weakly coupled dots.  In particular, this method is well-suited to the analysis of solid state quantum-dot-based qubits, and we consider the case of a double quantum dot (DQD) singlet-triplet qubit.  Past analyses have used techniques which are either substantially restricted in the regimes they can be used, or device specific and unsuited to exploration of a large regions of parameter space.  We analyze how the DQD exchange energy, which is central to the operation of qubit rotation gates, depends on a generic set of system parameters including magnetic field, DQD detuning, dot size, and dot separation.  We discuss the implications of these results to the construction of real devices.  We provide a benchmark of the CI by directly comparing results from the CI method with exact results for two electrons in a single parabolic potential (dot).  

%Understanding the DQD system is crucial to the construction of such qubits, and 
%This analysis is important for high-level guidance of experimental devices as well as optimizing over design parameters.

%TODO - remove some duplication & probably need to rework anyway
%Achieving low-error, exchange-interaction operations in quantum dots for quantum computing imposes simultaneous requirements on the exchange energy's dependence on applied bias.  A double quantum dot (DQD) qubit, approximated with a quartic potential, is solved using a full configuration interaction method.  We show that regimes can be found in which (1) the exchange energy $J$ is relatively insensitive to fluctuations in the bias voltage between the dots for positive and negative values of $J$ simultaneously, and (2) the magnitude of $J$ corresponds to qubit rotation times that are much greater than electronics jitter.  Identifying such regimes may prove valuable for the construction and operation of quantum gates that are robust to charge fluctuations.  Implications to a dynamically decoupled $z$-rotation gate is discussed.
\end{abstract}

%\pacs{todo}
\maketitle

%Main points
% - CI as powerful method for exploring parameter space
% - this is ptic important for noise considerations; finding ``flat'' areas
%     in this space would be beneficial
% - also important as semi-quantitative tool, exceeding HL, HM approximations
%     thus giving better order of magnitude estimates for the exchange energy
% - Below we highlight the qualitative features of quite general DQD potentials, some of which are not captured by more approximate methods.

\section{Introduction\label{secIntro}}
%TODO -- more method focused -- Rick?
%The fundamental unit of quantum computing is called a quantum bit, or qubit.  A qubit can be defined as an effective two-level quantum mechanical system with a long coherence time that can also be controlled and measured.  Thus, the state space of a qubit, however complex the actual system, can be mapped onto that of a single spin-$\frac{1}{2}$.  Operations on the qubit correspond to rotations of the spin-$\frac{1}{2}$ qubit, and are thus referred to as ``qubit rotations''.  Architectures have been proposed for implementing a qubit using a wide range of physical systems.  These include superconducting flux\cite{YuFluxQubit_2002,ChiorescuFluxQubit_2003} and charge\cite{NakamuraCooperBox_1999,Pashkin_2003} qubits, donor spin\cite{KaneNature_1998,HollenbergDonors_2004} and quantum dot\cite{BurkardLossDivincenzo_1999,PettaScience_2005,TaylorNatPhys_2005,VanDerWiel_spinQubits_2006} qubits in semiconductors, ion-trap qubits,\cite{PoyatosIonTrapGate_1997,PoyatosHotIons_1998,KielpinskiIonComputer_2002} and topological qubits which use non-abelian fractional quantum hall states.\cite{NayakNonAbelian_2008}  In many of the solid-state architectures (\emph{e.g.}~Loss-DiVincenzo,\cite{BurkardLossDivincenzo_1999} Kane,\cite{KaneNature_1998} and singlet-triplet\cite{PettaScience_2005}), a tunable exchange interaction between localized electrons is utilized to perform qubit operations.  The exchange interaction causes a splitting between quantum states (with different spin) called the \emph{exchange energy}, which we denote $J$.

The understanding and analysis of devices consisting of multiple coupled quantum dots is becoming increasingly important.  This is particularly evident in the area of quantum computing, where there are proposals for solid state qubits built from double quantum dots.  Such qubits must interact with each other (at least pairwise) to facilitate the two-qubit gates necessary for universal quantum computation.  Several standard approaches are used to simulate multi-dot systems, ranging from classical electronics solvers, which are ideal for large structures with many-electron quantum dots, to exact diagonalization techniques such as the full configuration interaction method, which are ideal for small systems of few-electron dots.  In this work, we are concerned with the latter, and specifically with understanding the physics underlying a general system of quantum dots.  We present here a full configuration interaction (CI) method which uses a Gaussian basis and is well-suited to the many-quantum-dot devices used to construct solid state qubits.  The method differs in small but significant ways from the approaches used in previous work, and is applied to gain insight into the double quantum dot (DQD) structures relevant to quantum computing.  It is capable of extending not only quantitatively upon prior work limited to more approximate methods, but also \emph{qualitatively} by capturing new aspects of DQD behavior.  In Ref.~\onlinecite{NielsenExchangePaper}, for instance, the (0,2)-occupation regime is studied in detail with respect to noise robustness in DQD qubits.  The DQD results we present here are more than just a slight improvement on previous, more approximate calculations; they show novel behavior that has not been explored to any great extent.  Before studying the DQD, we provide a direct comparison between the CI and a known exact result to help assess the degree of accuracy achieved by the method.  

In section \ref{secCalc} we describe the method in detail, highlighting its differences from similar past approaches.  In section \ref{secSingleDot}, we benchmark the CI using an exactly solvable model: two electrons in a 2D parabolic potential.  We consider there the convergence of the method with basis size, and find that with a relatively small basis quite accurate 2-electron energies are obtained.  In section \ref{secDoubleDot} the CI is applied to a singlet-triplet qubit.\cite{PettaScience_2005,TaylorDQDs_2006}   We study the exchange energy $J$ as a function of parameters which specify the electrostatic potential's shape and the magnetic field.  Past studies of the exchange energy in double-dot structures have been limited to either more approximate methods (\emph{e.g.}~Heitler London (HL) and Hund Mulliken (HM)) or to sophisticated finite-element numerical simulations of specific device structures.  Here we are able to give a more comprehensive picture of DQD physics since the CI is operable in regimes inaccessible to HL and HM approaches while still maintaining the computational speed necessary for extensive traversal of parameters space.

%The CI method used is more general than Heitler London (HL), Hund Mulliken (HM), and Hubbard model approaches, and requires significantly less computation resources than techniques which discretize the system using a mesh (\emph{e.g.}, finite-element methods).

%We use the CI in an explanation of the general trends in the exchange energy as a function of multiple electrostatic potential parameters and the magnetic field strength.

%??
%The computational basis (\emph{i.e.}~the levels of the effective 2-state system) consists of the two-electron singlet (spin $S=0$) and $S_z=0$ triplet ($S=1$) states of lowest energy.  The exchange energy $J$ is the splitting between these two states.

\section{Calculation\label{secCalc}}
%We compute the exchange energy of a DQD with two electrons using a full configuration interaction (CI) approach which constructs its single-particle basis out of Gaussian functions.  The ability to compute Hamiltonian matrix element analytically in a Gaussian basis results in a performance advantage over techniques requiring a large mesh. 
We now describe the configuration interaction (CI) method used to solve a $n$-electron effective mass Hamiltonian of the form 
\begin{equation}
%Many-particle Hamiltonian
\Hmb = \sum_i^n \Hop_i + \sum_{i < j}\frac{e^2}{\kappa |r_i-r_j|} \,,
\label{eqMBHam}
\end{equation} 
where $\kappa$ is an effective dielectric constant, and the single particle Hamiltonian for the $i^{\mathrm{th}}$ particle is given by 
\begin{equation}
%Single particle Hamiltonian
\Hop_i = \frac{(\vc{p}-e\vc{A})^2}{2m^*} + V(\vc{r})+ \frac{e}{m^*}\vc{S} \cdot \vc{B} \,,
\label{eq1PHam}
\end{equation}
where $\vc{r}$ and $\vc{p}$ are the position and momentum, respectively, of the $i^{\mathrm{th}}$ electron.  $V$ is the single-particle potential function, and $m^*$ is the effective mass (generally a tensor).  A vector potential $\vec{A}$ determines the magnetic field $\vc{B} = \vc{\nabla} \times \vc{A}$, which we restrict to be constant and along the $z$-direction: $\vc{B}=B\hat{z}$.

Overall, the approach follows the standard CI prescription\cite{CImethodBook} of finding a single-particle basis, building many-particle states from this basis, and exactly diagonalizing the Hamiltonian in the resulting many-particle basis.  In the present approach the single particle basis states are linear combinations of $n_G$ s-type Gaussian functions (in real space) of the form
\begin{equation}
%gaussian function (not nec. isotropic)
\begin{array}{c}
g(x,y,z) = N e^{-\alpha_x(x-x_0)^2}e^{-\alpha_y(y-y_0)^2} \\
\hspace{2cm} \times\, e^{-\alpha_z(z-z_0)^2} e^{\frac{ieB}{2\hbar}\left(y_0x-x_0y\right)}\,,
\end{array}
\end{equation}
where $N$ is a normalization factor, $B$ is the magnetic field, and $\vec{r}_0 = (x_0,y_0,z_0)$ and $\vec{\alpha}=(\alpha_x,\alpha_y,\alpha_z)$ are the position and exponential coefficient of the Gaussian function, respectively.  The aim in using such functions is to reduce the basis size needed to accurately model a system while allowing nearly all of the Hamiltonian matrix elements to be computed analytically, which results in a substantial performance advantage over approaches which discretize the system using a dense real-space mesh. Derivations and final expressions for the relevant matrix elements are given in Appendix \ref{appGaussianMatrixEls}.  Once the locations and exponential factors of the Gaussians are set (see below), an orthogonal single-particle basis is found by diagonalizing the single-particle Hamiltonian (Eq.~\ref{eq1PHam}) in the non-orthogonal basis of Gaussian functions (a generalized eigenvalue problem).  All the two-particle Slater determinant states from the single-particle basis are formed, and used as a many-body basis.  The full Hamiltonian (Eq.~\ref{eqMBHam}) is diagonalized in this basis, resulting in the system energies and eigenstates.  The set of Gaussian functions is optimized over a subset of fixed number and arrangement as follows.  The number, arrangement, and initial locations and widths of the Gaussian functions are input at the beginning of each run.  The set of Gaussian functions is then optimized by changing their locations and widths (while maintaining the same overall arrangement) to minimize the many-body ground state energy of a given symmetry sector of the Hamiltonian.  When states with different symmetries are of interest, the minimization is done multiple times (or, equivalently, the Hamiltonian matrix could be block-diagonalized and basis optimization performed within each block separately).  In the case of exchange calculations, where the lowest energy singlet and unpolarized triplet states are required, the singlet and unpolarized triplet energies are computed using separate basis optimizations.  Details of this optimization are given in Appendix \ref{appOptimizeGaussians}.

\section{Single Parabolic Dot\label{secSingleDot}}
We first consider an important test case: a single parabolic quantum dot in two dimensions.  With two interacting electrons, the system energies can be found exactly (the computation can be done analytically for certain dot confinement energies and numerically for any set of parameters using a one-dimensional Schrodinger solver).\cite{Taut_2elecExact_2009}  Comparison with this exact solution provides a useful benchmark for the CI, and we are able to analyze how well a Gaussian basis is able to capture a strongly-correlated two-electron state.  By varying the basis size, information on the convergence of the CI is obtained.  As a side remark, we note that the CI will always find the correct single-electron ground state energy since one of the Gaussian basis elements is chosen to be exactly this solution. %Thus, in a two electron double-dot system, it is much harder to correctly reproduce the (0,2) state then the (1,1) state.

%Thus with a single electron the 2D harmonic oscillator can of course be solved analytically. and since one of the basis elements used by the CI is the exact single-particle ground state, the CI will always be able to reproduce the groud state of the singly-occupied case.

%\subsection{Model}
Consider a single parabolic dot, given by the potential
\begin{equation}
V(\vc{r}) = \frac{1}{2}m^*\omega_0^2 r^2 \label{eqSingleDot}
\end{equation}
where $m^*$ is the effective mass and $\hbar\omega_0$ is the confinement energy of the dot.  We insert this potential into Eq.~\ref{eq1PHam} and then consider the full Hamiltonian given by Eq.~\ref{eqMBHam}, where $n=2$ is the number of electrons (there is no Coulomb repulsion term when $n=1$).  We use GaAs material parameters: $\kappa = 12.9$, and $m^* = 0.067\,m_e$.

\begin{figure}[h]
\begin{center}
\hspace{-2cm}
\parbox{2.7in}{\includegraphics[width=1.9in,angle=270]{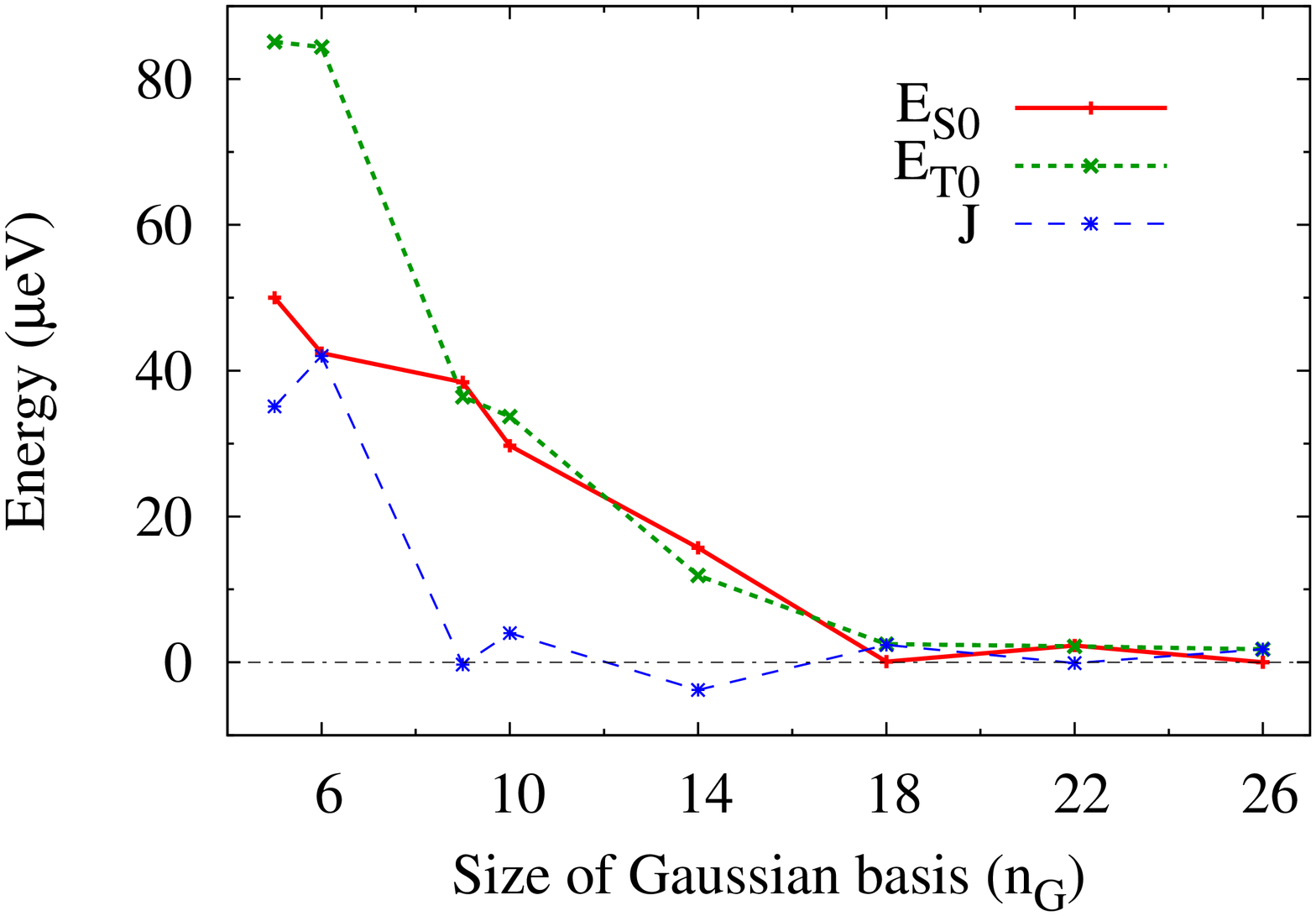}}
\parbox{0.2in}{
\begin{tabular}{|c|c|}
\hline
$n_G$ & Fig.\ref{figSpatialArrangements} \\
\hline
5 & A \\
6 & A* \\
9 & B \\
10 & B* \\
14 & C* \\
18 & B+B \\
22 & B+C \\
26 & B+D \\
\hline
\end{tabular} \vspace{0.2cm}}
\caption{The CI-computed lowest lying singlet ($E_{S0}$) and triplet ($E_{T0}$) energies \emph{relative to the exact value} as a function of basis size.  The energy,   Their difference, the exchange energy $J=E_{T0}-E_{S0}$ is also given relative to the exact value.  The spatial arrangement of the basis elements for each size is given in the table to the right, which gives a letter A-D of a spatial plot in Fig.\ref{figSpatialArrangements}.  A trailing Astrix's (*) indicates that there are two basis elements lying on top of one another at the center of the dot, and the sum (+) of two letters indicates that the elements of the referred to spatial plots are combined (elements at the same position have different exponential factors).  Dot parameters $\hbar\omega_0=3\meV$ and $B=0$.\label{figCompareWithExact}}
\end{center}
\end{figure}

\begin{figure}[h]
\begin{center}
\includegraphics[width=1.2in,angle=270]{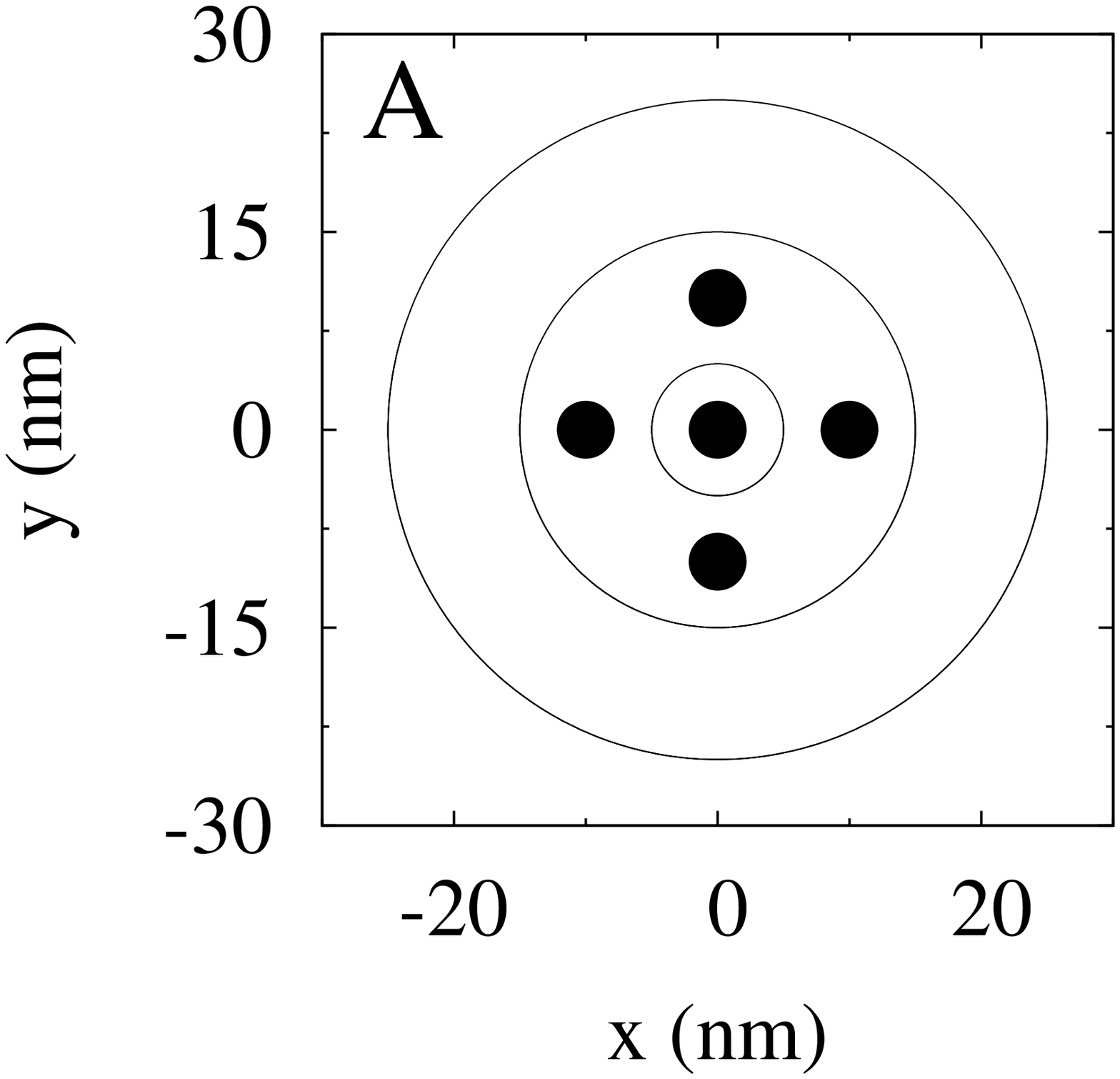} \hspace{-1cm}
\includegraphics[width=1.2in,angle=270]{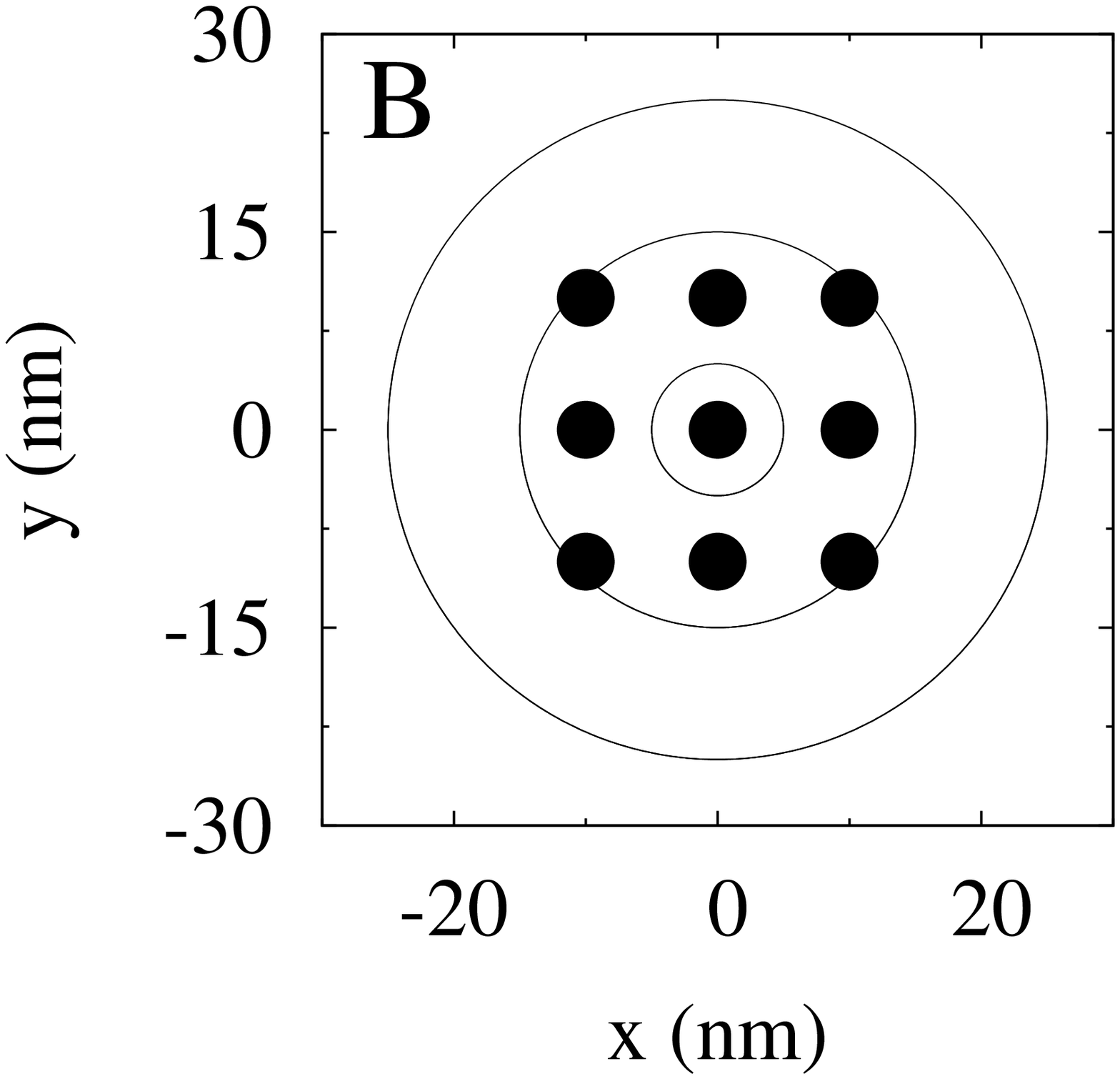}
\includegraphics[width=1.2in,angle=270]{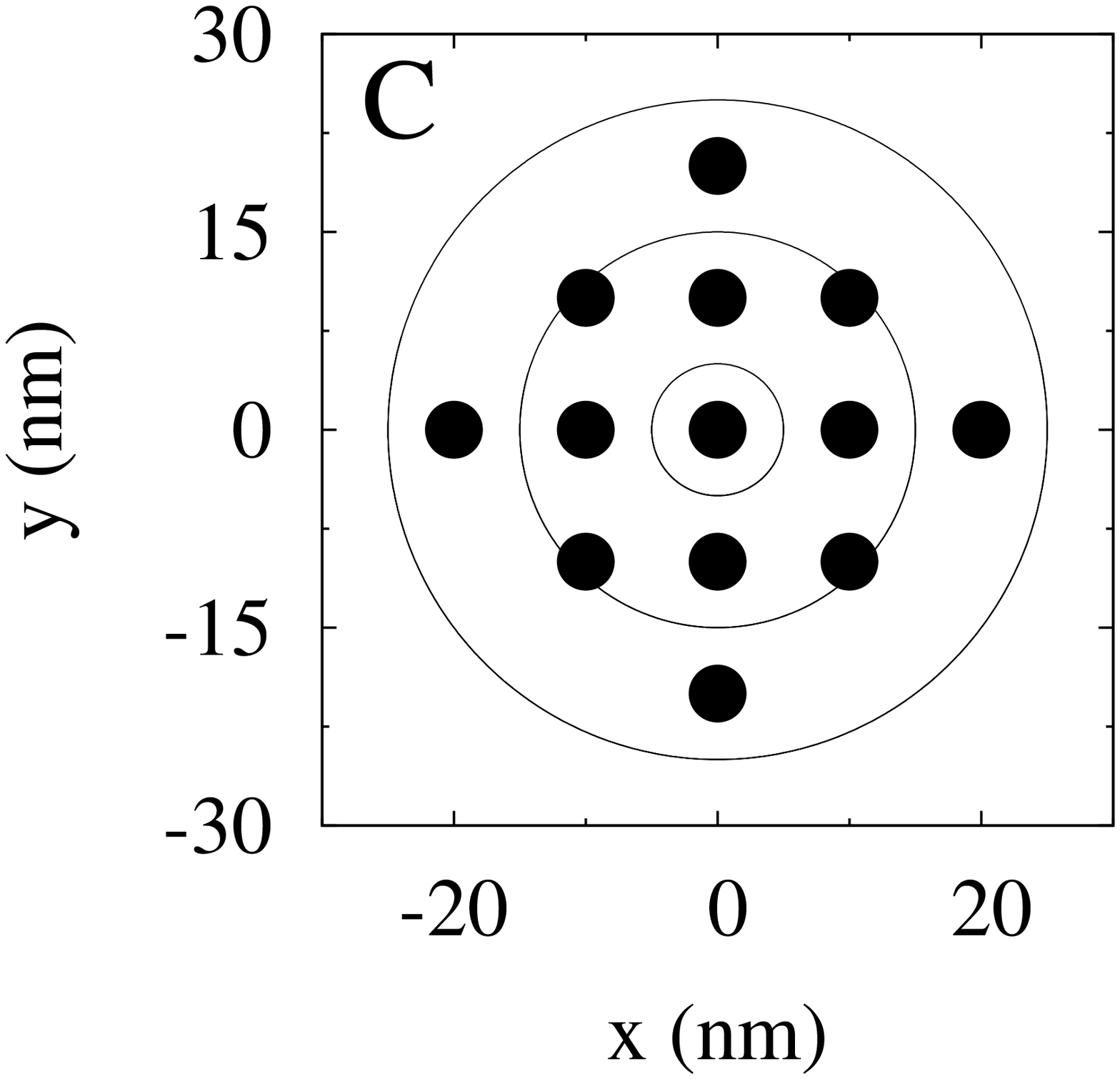} \hspace{-1cm}
\includegraphics[width=1.2in,angle=270]{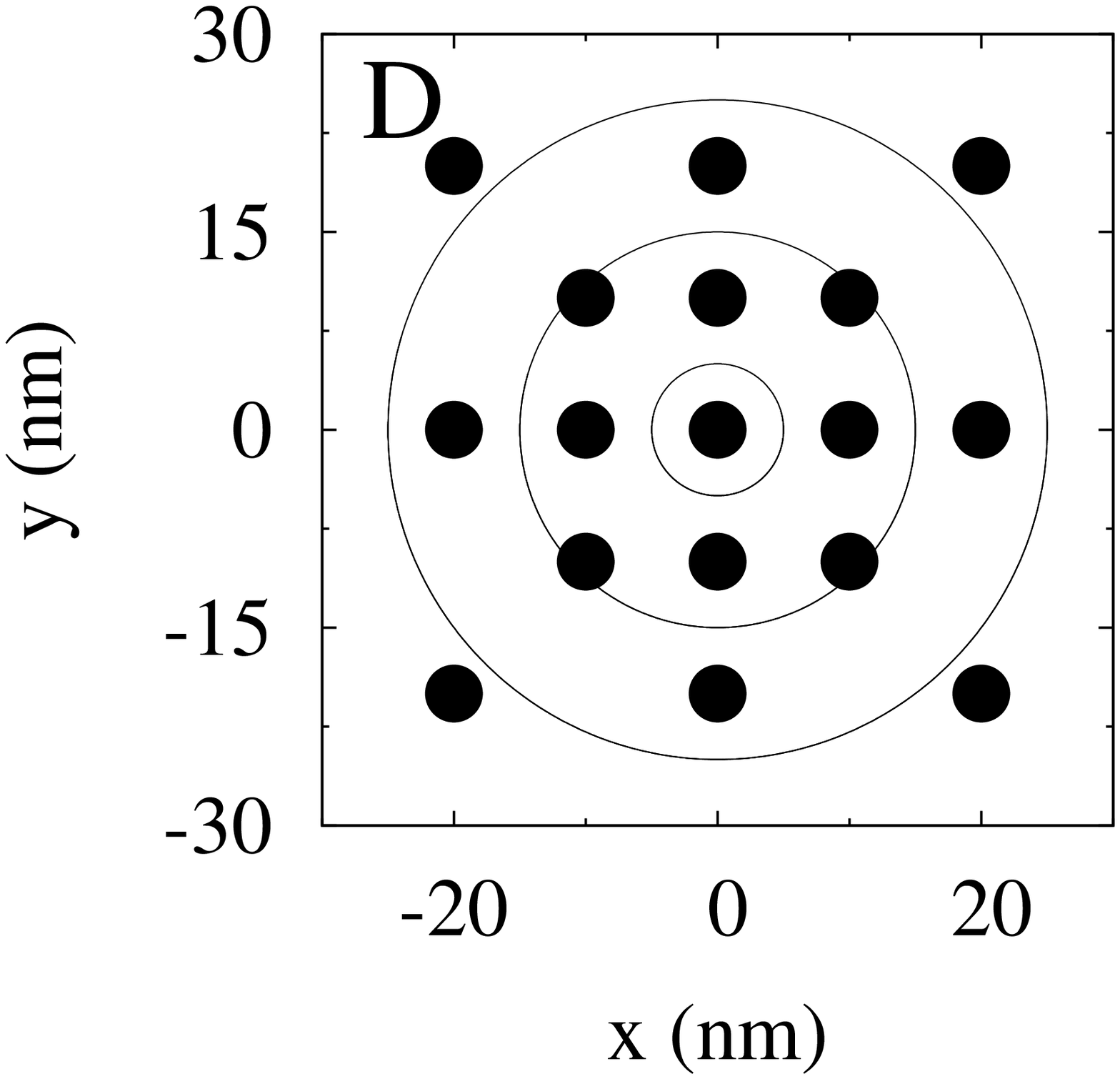}
\caption{Spatial arrangements of Gaussian basis elements in a single quantum dot.  Each solid black circle represents at least one basis element.  There can be more than one element at a given point (circle) as specified by the table in Fig.~\ref{figCompareWithExact}.  Note that these diagrams only describe the relative arrangement of the Gaussian centers; the ``grid spacing'' of the centers is optimized.  \label{figSpatialArrangements}}
\end{center}
\end{figure}

%HERE
%\subsection{Results}
We solve the full Hamiltonian exactly using the method in Ref.~\onlinecite{Taut_2elecExact_2009} to reduce the problem to an ordinary (one-dimensional) Schrodinger equation by switching to center of mass coordinates.  We then solve this equation using the technique prescribed in Ref.~\onlinecite{Sudiarta_SchrodingerSolver_2007}.  The lowest singlet (total spin $S=0$) and unpolarized triplet ($S=1$, $S_z=0$) energies obtained by the CI relative to the exact solution are shown in Fig.~\ref{figCompareWithExact} as a function of basis size.  As size does not uniquely specify a basis, we give the spatial arrangement of the basis elements used in Fig.~\ref{figSpatialArrangements} which is referenced by the table in in Fig.~\ref{figCompareWithExact}.  Overall, we find for a range of dot confinement $\omega_0$ and magnetic field $B$ that the CI energies converge to within $\approx 0.5\%$ of their value for basis sizes around 10, and to within $\approx 0.05\%$ for basis sizes around 20.  Thus, for the dot parameters of Fig.~\ref{figCompareWithExact} where the energies are of order $10\meV$, convergence is obtained to within $50$ and $5\mueV$ for roughly 10 and 20 basis elements, respectively.  This assumes a good basis arrangement (cf.~Fig.~\ref{figSpatialArrangements}), as a poor choice of where to place the basis elements (\emph{e.g.}~all in a single line) will clearly not produce converged values even for large basis sizes.

\section{Double Parabolic Quantum Dot\label{secDoubleDot}}

We proceed to consider a double quantum dot (DQD) system, which cannot be solved or substantially simplified analytically.  It is directly relevant to quantum computing as it can be used to implement the fundamental unit for quantum computation, a quantum bit (qubit).  A qubit can be defined as an effective two-level quantum mechanical system with a long coherence time that can also be controlled and measured.  Thus, the state space of a qubit, however complex the actual system, can be mapped onto that of a single spin-$\frac{1}{2}$.  Operations on the qubit correspond to rotations of the spin-$\frac{1}{2}$ qubit, and are thus referred to as ``qubit rotations''.  While architectures have been proposed which implement qubits using a wide range of physical systems, many of the solid-state approaches (\emph{e.g.}~Loss-DiVincenzo,\cite{BurkardLossDivincenzo_1999} Kane,\cite{KaneNature_1998} and singlet-triplet\cite{PettaScience_2005}), use a tunable exchange interaction between localized electrons to perform qubit operations.  The exchange interaction causes a splitting between quantum states (with different spin) called the \emph{exchange energy}, which we denote $J$.  

We focus here on a singlet-triplet qubit where two electrons are localized in a double quantum dot.  The dots are laterally spaced near a material interface, and lithographically defined gates in a plane parallel to the interface create and control the electrostatic potential forming the quasi-two-dimensional dots.  We assume the dot potentials are parabolic, which is makes the system particularly suited to analysis by our Gaussian-basis CI method.  We use this example to both illustrate the capabilities of the method described in section \ref{secCalc} and to report, using GaAs material parameters, the trends of $J$ as function of the DQD electrostatic potential and magnetic field, which we believe is helpful for the understanding and design of such devices. %check this sentence

The results we present can only be treated semi-quantitatively for several reasons.  First, the exact form of the potential is unknown and the problem is only approximately two-dimensional.  In an actual device, the dots will not be perfectly parabolic and the electrons are confined to a finite width in the direction perpendicular to the surface.  Second, we use nine Gaussian basis elements per dot, and even though this results in energy levels converged to approximately $0.5\%$ of their value, the exchange energy can have substantial fractional error when it's absolute value is small (below $100\mueV$ for our purposes).  Even so, our results give a more accurate qualitative and semi-quantitative picture than previous variational approaches, and are sufficient to resolve the features of interest.  In cases where additional accuracy is required, more powerful CI and iterative techniques (\emph{e.g.}~Poisson-Schr\"{o}dinger solvers) can be used.\cite{Abolfath_HFCI_2006,StopaImmunity_2008}   Such techniques, however, are more computationally intensive and are best used to model realistic device structures rather than provide qualitative insight.

While variations of the CI method with Hartree-Fock\cite{Abolfath_HFCI_2006} or molecular orbitals\cite{HuHilbertSpace_2000,Kettle_MODonor_2006,Zhang_ellipticExchange_2007,Zhang_ellipticalNonMono_2009} have been used in the past, to our knowledge this is the first application of a CI method to a biased DQD capable of modeling both weakly and strongly coupled regimes.  (The CI method in Ref.~\onlinecite{HuHilbertSpace_2000} is used to analyze an unbiased DQD, and is also restricted to the weakly coupled case.)

\subsection{Model\label{ssecModel}}
\begin{figure}[h]
\begin{center}
\includegraphics[width=2.2in,angle=270]{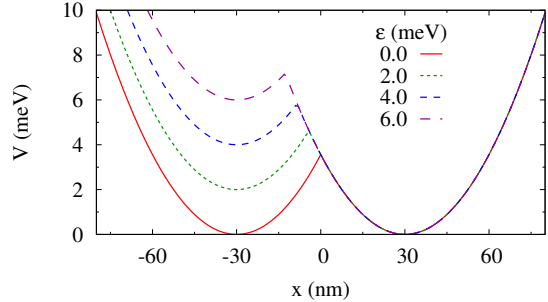}
\caption{DQD potential along x-axis, $V(x,0)$, for $L=30\nm$ and $E_0=3.0\meV$.  $V$ is the minimum of parabolas centered at $x = \pm L$ with curvature proportional to $E_0^2$. \label{figPotential}}
\end{center}
\end{figure}

A DQD qubit is considered within the effective mass approximation, and the electrostatic potential is approximated as the minimum of two parabolic dots, so that $V$ has the form
\begin{equation}
\begin{array}{c}
V(x,y,z) = \frac{1}{2}\left[ m^*_x\omega_x^2 \min\left( (x-L)^2 + \epsilon, (x+L)^2 \right)\right. \\
\vspace{-0.1cm}\\
\left. + m^*_y \omega_y^2 y^2 + m^*_z \omega_z^2 z^2\right]\,.
\end{array} \label{eqPot}
\end{equation}
This potential is parametrized in terms of $\epsilon$, $L$, and $\vc{\omega}$.  It is readily seen from Eq.~\ref{eqPot} that $\epsilon$ is the energy difference between the left and right minima (the inter-dot bias), $L$ is half the distance between the minima, and $\vec{E}_0\equiv \hbar\vec{\omega}$ is the confinement energy of the dots along the coordinate axes. In the two-dimensional case, the potential is given by Eq.~\ref{eqPot} with $z$ set to zero.  Cuts of the potential along the $x$-axis for different $\epsilon$ are shown in Fig.~\ref{figPotential}.  We use GaAs material parameters, $m^* = 0.067\,m_e$ and $\kappa = 12.9$ unless otherwise noted.

%We consider two-dimensional systems with material parameters for gallium arsenide in an AlGaAs well and silicon near a Si/SiO$_2$ interface.   and in the silicon case $(m^*_x,m^*_y,m^*_z) = (0.19,0.19,0.98)\,m_e$ and $\kappa = (\kappa_{Si} + \kappa_{SiO_2})/2 = 8.0$ (accounting for the image charge in the oxide\cite{LiDasSarmaExchange_2009}).  In the case of silicon, we assume that the valley-splitting is large compared to the energies we consider so that a single-valley treatment is valid.% (the expected valley splitting gap is $\approx 100\,\mu\mbox{eV}$\cite{valleySplittingNumber}).

%Since the electrostatic potential is created by lithographic gates, $V(\vc{r})$ (and thus the exchange energy) can be changed by varying the gate voltages.  Electrostatic control of $J$ has the advantage of allowing fast qubit operations, but also provides an avenue through which charge noise couples to exchange noise.

In a real device, fabrication technique and gate voltages can effectively modify each of the parameters $\epsilon$, $L$, and $\vec{E_0}$; the device design and voltage sequence will determine exactly how these parameters vary during device operation.

%TODO: Add sentence ``we have in mind a device...''??

%     In general, changing the gate voltages in order to vary the exchange interaction will alter the shape of the electrostatic potential in a way which corresponds to varying a combination of $\epsilon$, $L$, and $E_0$, and $\omega_y/\omega_x$.  We have in mind a device in which gate operations are carried out by varying $\epsilon$ independently of the other variables (in principle a control sequence could be engineered to change any of the electrostatic potential parameters independently of the others).  

% We have in mind a device in which gate operations are carried out by varying $\epsilon$ independently of the other variables. In principle adjacent gate voltages can provide compensation to effectively change any parameter independently.

%TODO: Charge sector definition??

\subsection{Exchange Energy Results\label{ssecResults}}
\subsubsection{General Features}
Here we report, using GaAs material paramters, the general trends of $J$ as function of the DQD electrostatic potential and magnetic field.  We have in mind a system that varies $\epsilon$ to perform qubit rotations, and so focus on the $\epsilon$ dependence of $J$.  Qualitatively similar behavior is seen in the Si/SiO$_2$ system, as will be shown in section \ref{ssecSiComparison}.

% - 1 single J vs. eps curve with N2 on right y-axis, label intervals I,II,III {figJvsEpsTypical}
%    show slope == 1?
\begin{figure}[h]
\begin{center}
\includegraphics[width=2.2in,angle=270]{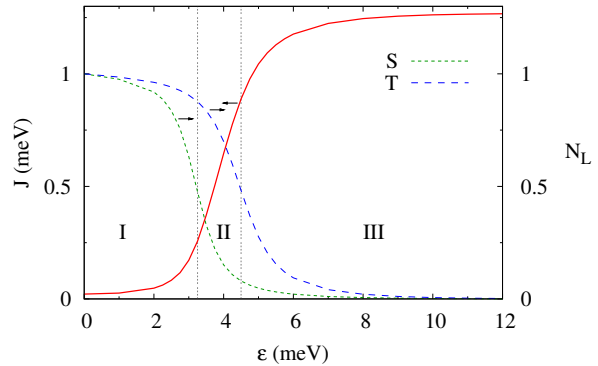}
\caption{Typical zero magnetic field exchange energy (solid line) as a function of dot bias.  Dashed lines show the electron occupation of the left dot for the singlet and triplet state.  The domain is divided into three intervals (I,II,III) based on the the charge configuration of the singlet and triplet states (see text).  Electrostatic parameters are $L=30\nm$, $E_0=3\meV$. 
%CI parameters are $n_G=39$, $n=70$.  
 \label{figJvsEpsTypical}}
\end{center}
\end{figure}
%vertical stacking - 2D e-density plots?

% - several J vs. eps curves for variable L; J vs. L for several eps (fix E0,B=0) {figJvsEpsL}
\begin{figure}[b]
\begin{center}
\includegraphics[width=2.2in,angle=270]{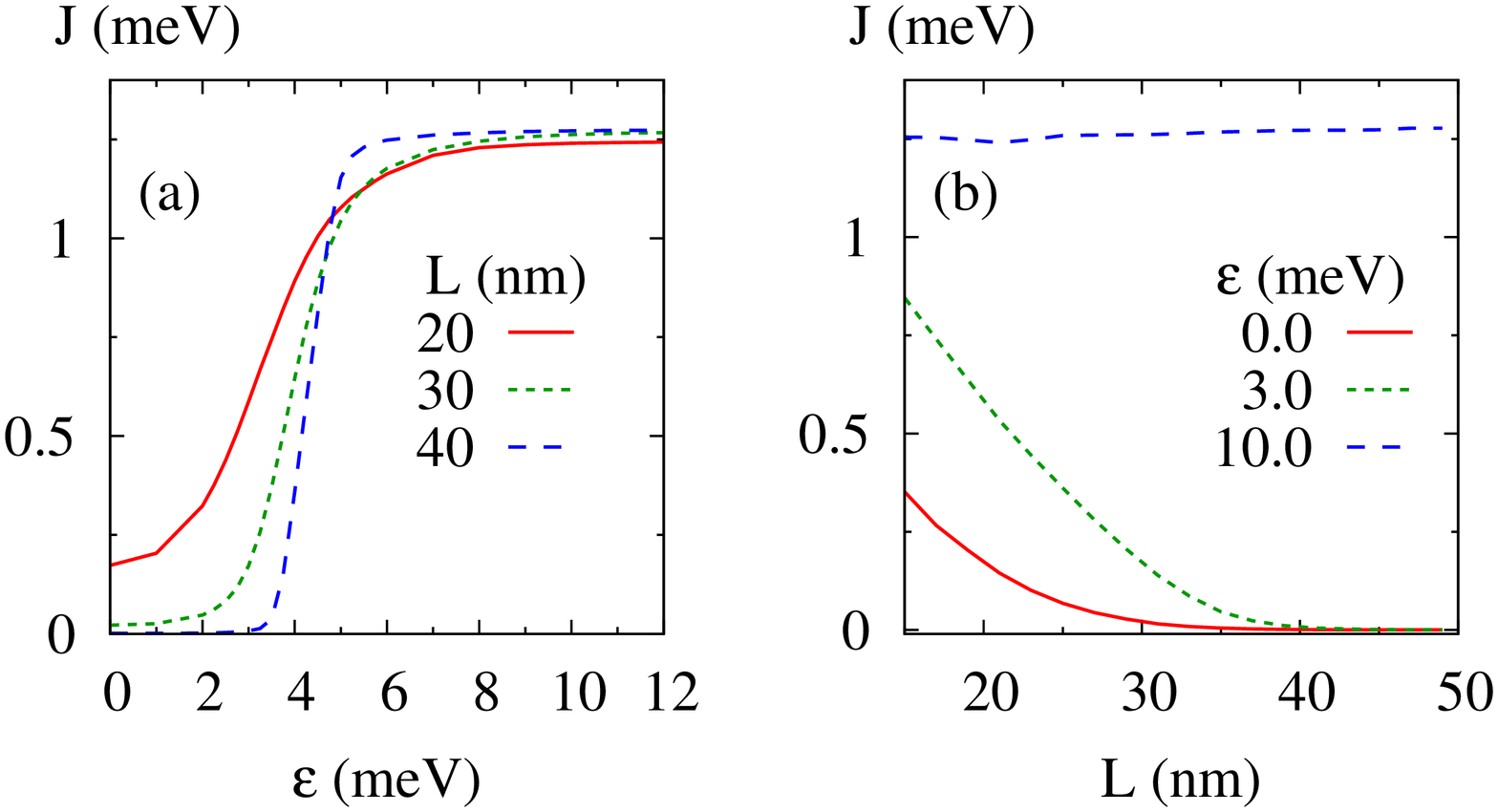}
\includegraphics[width=2.2in,angle=270]{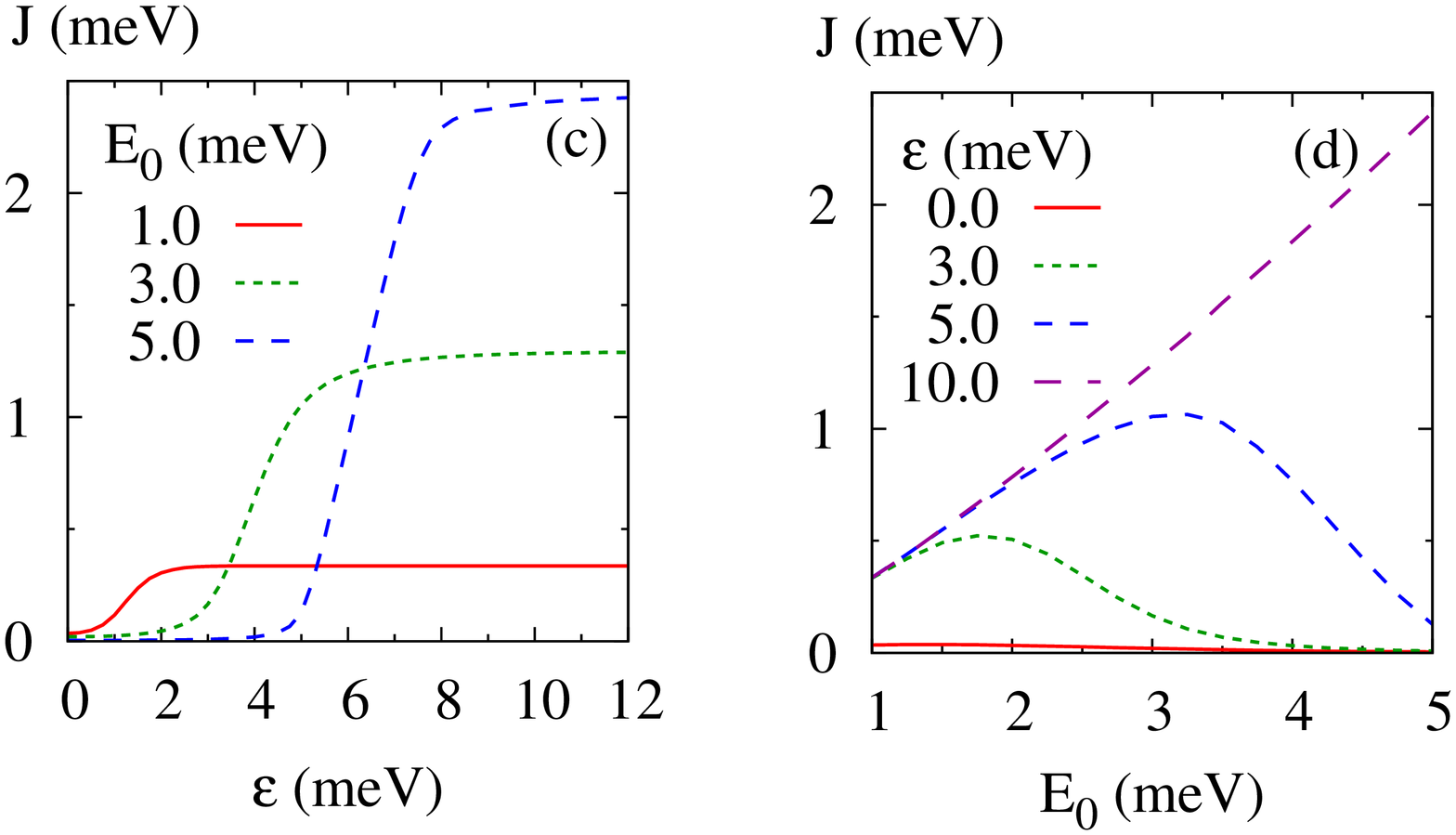}
\caption{Zero-field ($B=0$) behavior of the exchange energy when varying inter-dot separation ($2L$) and confinement energy ($E_0$).  In (a) $J$ vs.~$\epsilon$ curves become sharper steps and $J$ in the low-$\epsilon$ region (I) decreases as $L$ increases.  This is illustrated by $J$ vs.~$L$ curves in (b), which decrease at low $\epsilon$ and become flat at larger $\epsilon$, when the DQD is in the (0,2) charge sector.  In (c) it can be seen that with increasing $E_0$, $J$ vs.~$\epsilon$ curves rise at larger $\epsilon$ and flatten out at larger $J$.  Plots of $J$ vs.~$E_0$, shown in (d), increase until $E_0$ the DQD transitions to the (1,1) charge sector.  In (a) and (b), $E_0$ is fixed at $3\meV$, and in (c) and (d), $L=30$.
 %CI parameters are $n_G=39$, $n=70$.  
 \label{figJvsEpsLE0}}
\end{center}
\end{figure}

%TODO - check exponential dependence of J vs.L 
%     - show slope == 1 in figure??  if not change text below
First we consider the case of zero magnetic field.  The shape of a typical $J$ vs.~$\epsilon$ curve when $B=0$ is shown in Fig.~\ref{figJvsEpsTypical} (for $L=30\nm$ and $E_0=3\meV$).  The $\epsilon$-axis can be divided into three regions (marked in Fig.~\ref{figJvsEpsTypical}) corresponding to different charge character of the singlet and triplet states:  (I) low $\epsilon$, where both singlet and triplet are in the (1,1) charge sector, (II) intermediate $\epsilon$, where the singlet takes on (0,2) character and the triplet remains a (1,1) state, and (III) large $\epsilon$, where both states are in the (0,2) charge sector.  In (I), $J$ is due to the difference in Coulomb energy between spatially symmetric and antisymmetric compositions of single-dot wavefunctions, and scales exponentially with the inter-dot spacing (see Fig.~\ref{figJvsEpsLE0}).  Note that at $\epsilon=0$, we have $dJ/d\epsilon=0$ exactly, due to the symmetry in $\epsilon$.  In (II), $J$ increases rapidly because only the triplet is penalized for having an electron in the left dot.  %The slope of the linear portion in the center of (II) is therefore centered around one, as shown.
  The curvature at the beginning and end of the linear portion is due to the inter-dot tunnel coupling (which increases with decreasing $E_0$ and/or $L$).  Larger tunnel coupling results in a more gradual transition between (II) and the other two regions, as seen in Fig.~\ref{figJvsEpsLE0}.  The exchange energy levels off in (III) since $\epsilon$ does not affect the shape of the dot holding both of the electrons.  The derivative $dJ/d\epsilon$ can be made arbitrarily close to zero by increasing epsilon, and in Fig.~\ref{figJvsEpsTypical} decreases to $O(10^{-6})$ by $\epsilon=12\meV$.  The nearly constant value of $J$ in (III) is essentially the exchange energy of a doubly-occupied single dot with confinement energy $E_0$.  Figure \ref{figJvsEpsLE0} shows that $J$ is insensitive to $L$ in this region (since $L$ does not affect the dot shape), and that increasing $E_0$ raises the value of $J$.  Note also that the boundary between regions (II) and (III) moves to larger $\epsilon$ as $E_0$ is increases, so that at large enough $E_0$ the exchange energy actually decreases as $E_0$ increases (see Fig.~\ref{figJvsEpsLE0}c).
%, and Fig.~\ref{figSpatialDensity} displays snapshots of the singlet's electron density as it progresses from the (1,1) to (0,2) charge sector. 
% regardless of their spin configuration.

% - several J vs. B curves for variable eps (fix E0,L), lapbel intervals i,ii,iii - 2 plots for different E0??
% - several J vs. eps curves for variable B (fix E0,L) {figJvsB}
\begin{figure}[h]
\begin{center}
\includegraphics[width=2.2in,angle=270]{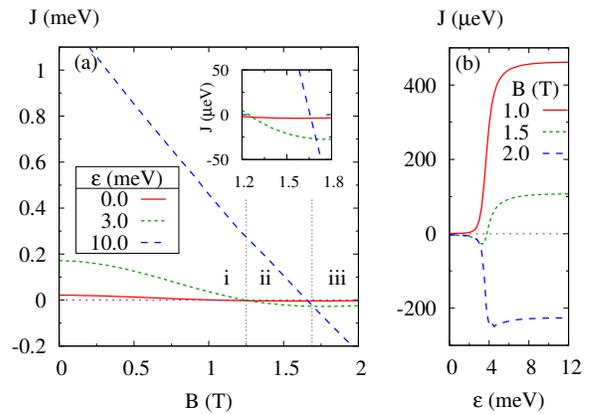}
\caption{Exchange energy as a function of magnetic field for $L=30\nm$ and $E_0=3\meV$.  In (a), $\epsilon=0,3,5\meV$ correspond to regions I, II, and III of $J$ vs.~$\epsilon$ curves respectively.  The three different sequences of the curves in (a), delimited by the intervals (i), (ii), and (iii), lead to three qualitatively different types of $J$ vs.~$\epsilon$ behavior.  This is shown in (b), where $B=1$, $1.4$, and $2\,\mathrm{T}$ fall in (i), (ii), and (iii).  The inset of (a) is a magnification around interval (ii).  %CI parameters are $n_G=39$, $n=70$.
\label{figJvsB}}
\end{center}
\end{figure}

% - proceed to show & explain B > 0 case
Now consider the case of finite magnetic field.  In a single dot with two electrons, $J$ oscillates as a function of $B$.\cite{WagnerSingleDot_1992,Dybalski_2005}  This is due to a competition between Coulomb and rotational energies which favors a state of increasingly higher angular momentum $l$. The ground state therefore alternates between an even-$l$ singlet and an odd-$l$ triplet state, resulting in an oscillatory $J$.  Increasing $B$ also increasingly confines the electrons and contributes to a reduction of $|J|$.  For a DQD with one electron in each dot, $J$ is also a decaying oscillatory function of $B$ which can be explained similarly.  This behavior is shown in Fig.~\ref{figJvsB}a for fixed $E_0$ and $L$, and for $\epsilon$ at representative values in intervals (I), (II), and (III) of Fig.~\ref{figJvsEpsTypical}.  We see that for a fixed $B$, the exchange energy can have three qualitatively different types of dependence on $\epsilon$:  (i) $J$ increases, (ii) $J$ decreases then increases, and (iii) $J$ decreases. These three scenarios divide the $B$-axis into the intervals (i)-(iii) shown in Fig.~\ref{figJvsB}a.  Curves of $J$ vs.~$\epsilon$ for $B$ in each of these intervals are shown in Fig.~\ref{figJvsB}b.  Most noteworthy is case (ii), in which $J$ possesses a local minimum.  This minimum exists because the $B$-field required to push $J < 0$ is less in the (1,1) than (0,2) charge sector, but the derivative $|dJ/dB|$ is larger in the (0,2) sector.\cite{StopaImmunity_2008}

\subsubsection{Convergence}
\begin{figure}
\begin{center}
\includegraphics[width=2.5in,angle=270]{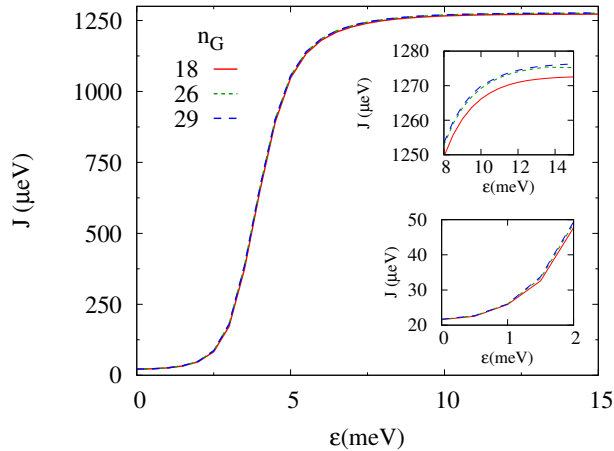}
\caption{Convergence of the CI method shown by comparing the exchange energy given by Gaussian bases with $n_G$ elements.  Parameters $E_0=3\meV$, $L=30\nm$, $\omega_x=\omega_y$, and $B=0$ are used, and represent the typical convergence behavior seen at other parameter values. Insets show zoomed views of the low- and high-$\epsilon$ regions of the main plot.\label{figConvergence}}
\end{center}
\end{figure}

As the number of basis elements increases, the eigenenergies converge to the exact spectrum of the many-body Hamiltonian.  Because of the basis optimization that is performed, a relatively small number of Gaussian basis elements ($n_g\approx 20$) result in energies converged on the scale of several micro electronvolts.  Figure \ref{figConvergence} shows an exchange vs.~bias curve using $n_g=18$, $26$, and $29$ basis elements.  The arrangements of these elements correspond to Fig.~\ref{figSpatialArrangements}b on each dot, Fig.~\ref{figSpatialArrangements}c on each dot, and Fig.~\ref{figSpatialArrangements}c on each dot with 3 elements between the dots, respectively.  As the insets of this figure reveal, the low-$\epsilon$ region converges more rapidly than the high-$\epsilon$ region.  This is expected since in the low-$\epsilon$ region the electrons are in separate dots and the multi-electron wavefunction is well approximated by linear combinations of the single-particle states in each dot, for which the Gaussian basis elements are excellent approximations.  The high-$\epsilon$ regime is that of a single doubly-occupied dot, which was considered in section \ref{secSingleDot}.  We note that even though the energies have only converged to tens of$\mueV$, the exchange energy has converged to approximately $5\mueV$ due to error cancellation.  Since the energies have magnitudes of order $10\meV$, the relative error is $\Delta E/E \approx 0.05\%$.

%Our results are still only semi-quantitative, since the exact form the the potential is unknown and the choice of basis can still have quantitatively significant effects on the exchange energy, especially in cases of small exchange interaction (it is difficult to accurately obtain a small number through the subtraction of two large ones).  

%Even with this limitation, our results give a more accurate qualitative and semi-quantitative picture than previous variational approaches.  In cases where additional accuracy is required, more powerful CI and iterative (\emph{e.g.}~Poisson-Schr\"{o}dinger solvers) techniques\cite{Abolfath_HFCI_2006,StopaImmunity_2008} can be used.  Such techniques, however, are much more computationally intensive since they require the calculation of matrix elements to be discretized over a mesh (matrix elements can be computed analytically for a Gaussian basis), and are best used to model realistic device structures rather than provide qualitative insight.

\subsubsection{Differences in Si/SiO$_2$\label{ssecSiComparison}}
\begin{figure}
\begin{center}
\includegraphics[width=2.2in,angle=270]{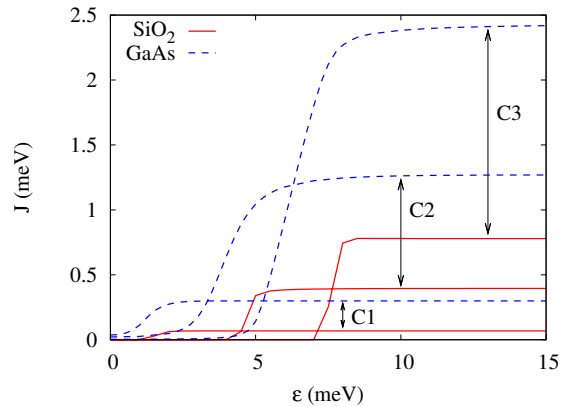}
\caption{Comparison of the exchange energy obtained using material parameters $m^*$ and $\kappa$ appropriate for GaAs and SiO$_2$ systems.  DQD potentials with three different curvatures C1, C2, and C3, are used to generate the three curves for each material.  These curvatures correspond to $E_0=1$, $3$, and $5\meV$ in the GaAs system, but $E_0$ will be different for the SiO$_2$ system as described in the text.  All DQD potentials use $L=30\nm$ and $B=0$. \label{figSiO2GaAsCompare}}
\end{center}
\end{figure}

%Since the CI treats the semiconductor within an effective mass approximation, it is straightforward to perform the calculation for materials other than GaAs so long as they have a single-valley.  
Without any modification to our method, results for a Si/SiO$_2$ system in the single-valley approximation can be obtained using appropriate material parameters ($(m^*_x,m^*_y,m^*_z) = (0.19,0.19,0.98)\,m_e$ and $\kappa = (\kappa_{Si} + \kappa_{SiO_2})/2 = 8.0$, which accounts for the image charge in the oxide\cite{LiDasSarmaExchange_2009}).  For the single-valley approximation to be valid, the splitting between states with different valley character must be large compared to the splitting between the ground and excited states of interest.  We note that because the definition of $E_0$ depends on the effective mass, the same quantum dot potential specified by $\vec{E}_0^{GaAs}$ in a GaAs system will be given by $\vec{E}_0^{Si}$ in a silicon system, where $E_{0i}^{Si} = E_{0i}^{GaAs}\sqrt{m_i^{GaAs}/m_i^{Si}}$.  Thus, the corresponding $E_0$ values for $E_0=1$, $3$, and $5\meV$ in GaAs are $E_0=0.594$, $1.782$, and $2.969\meV$ in a Si/SiO$_2$ system.  As expected, the exchange energy of an SiO$_2$ system shows qualitative behavior identical to GaAs systems.  Quantitatively, the larger effective mass and smaller dielectric constant together result in lower values of the exchange energy given the same DQD potential (see Fig.~\ref{figSiO2GaAsCompare}).  This comparison is somewhat artificial, however, since different devices and experimental parameters (\emph{e.g.}~gate voltages) would be required to create identical potentials in the different materials.  %TODO - explanation given scaling argument??

\subsubsection{Comparison with other approaches}
%Comparison of HL, HM, CI
Lastly, we compare the CI method with Heitler London (HL) and Hund Mulliken (HM) techniques which previous studies\cite{Mizel_3or4Dots_2004,Hada_SiDQDExchange_2004,HuFluctuation_2005,CaicedoOrtizDQDExchange_2006,VanDerWiel_spinQubits_2006,Hatano_fewElDQD_2008,LiDasSarmaExchange_2009} use to study DQD exchange energy.  The CI outlined here is more general than these methods from a variational standpoint:  The full CI is a variational method and the space of trial wavefunctions includes the HL and HM spaces as long as the number of Gaussian functions $n_G \ge 2$ (and is identical to HM for $n_G=2$).  Effective Hubbard models are usually less precise than HM, since they approximate the Coulomb interaction as short-ranged.  The small variational spaces of these methods restrict their applicability to qualitative results for \emph{weakly} coupled quantum dots,\cite{Calderon_HLValidity_2006,Pedersen_FailureStdApproaches_2007} though the addition of variational parameters to the basis improves their accuracy.\cite{Saraiva_HLReliability_2007,Zhang_2008,Dybalski_2005}

\begin{figure}[h]
\begin{center}
\includegraphics[width=2.2in,angle=270]{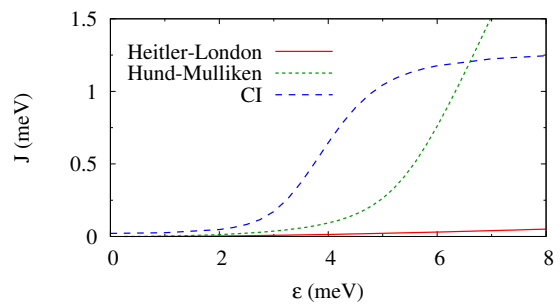}
\caption{Comparison of configuration interaction (CI), Hund-Mulliken(HM), and Heitler-London(HL) methods for $E_0=3\meV$, $L=30\nm$, and $B=0$.  Neither HL or HM methods can capture the flattening of the $J$ vs.~$\epsilon$ curve at large $\epsilon$. \label{figHMCompare}}
\end{center}
\end{figure}

Figure~\ref{figHMCompare} shows a comparison of the CI, HM and HL methods for the parameters of Fig.~\ref{figJvsEpsTypical}.  Neither the HL nor HM method can predict the large-$\epsilon$ flat since neither include triplet (0,2) states.  While the HM method captures the sharp rise in $J$ (since it includes a (0,2) singlet), the onset of the rise is at much larger $\epsilon$.  This is due to the tunnel barrier between the dots begin effectively larger because of the absence of basis elements which have support between the dots.  Said another way, in the HM case an electron can be centered in either the left or right dot but not between them, and therefore it requires a larger bias to push the electron out of the right dot.

\section{Conclusion} 
We have described a configuration interaction method which uses Gaussian functions as basis elements and applied it to two-electron single- and double-quantum-dot (DQD) systems, the latter functioning as a singlet-triplet qubit.  The increased accuracy and extended regimes of validity of the CI relative to HL and HM methods, coupled with its ability to explore parameter space much more rapidly than mesh-based solvers, make the method uniquely suited to analyze the exchange energy in DQDs.  We explain the general behavior of the exchange energy as a function of DQD parameters which correspond to inter-dot bias, inter-dot separation, dot size, and magnetic field.  In particular, we are able to analyze the transition from (1,1) to (0,2) charge sector, where more approximate methods fail, for both weakly and strongly coupled dots.  This reveals new qualitative features of the exchange curve which could be useful for suppressing the effects of charge noise in physical qubits.\cite{NielsenExchangePaper}  We focus on two-dimensional dots in a GaAs system, and by considering SiO$_2$ systems find that the qualitative features of the exchange energy are insensitive to typical changes in material parameters.  The semi-quantitative results given in this analysis can be used to guide the tuning of experimental DQD devices seeking to implement a singlet-triplet qubit.

%We have carried out configuration interaction calculations on a singlet-triplet DQD qubit in a single valley approximation. We demonstrate that for electrostatic potentials and constant magnetic fields, the exchange energy $J$ as a function of DQD bias voltage $\epsilon$ will have locally flat regions ($dJ/d\epsilon \approx 0$) both for positive and negative $J$ that are of order$\mueV$, compatible with realistic $1-10\ns$ gate times.  We discuss how these features can benefit a qubit's robustness to charge fluctuations, and how they might be used to realize a noise-protected $z$-rotation quantum gate. The CI method used offers more than just semi-quantitative improvement over more approximate schemes such as HL or HM (though one advantage of the CI over these methods is the ability to more accurately consider \emph{strongly} coupled dots).  Namely, it captures the flat region of the $J$ vs.~$\epsilon$ curve at large $\epsilon$ and the full transition from small- to large-$\epsilon$ regimes.

This work was supported by the Laboratory Directed Research and Development program at Sandia National Laboratories. Sandia is a multiprogram laboratory operated by Sandia Corporation, a Lockheed Martin Company, for the United States Department of Energy’s National Nuclear Security Administration under Contract DE-AC04-94AL85000.

\appendix

\begin{widetext}
\section{Matrix Elements between s-type Gaussian functions \label{appGaussianMatrixEls}}

\subsection{1P Gaussian matrix elements \label{sec1PMatrixEls}}

We derive here the matrix elements for single-particle operators which are piecewise polynomial in the components of particle position $\vc{r}$ and momentum $\vc{p}$.  Since the basis functions are Gaussian, this reduces to the case of operators piecewise polynomial in the components of $\vc{r}$ ($\vc{p} = -i\hbar\nabla$ brings down powers of $\vc{r}$ from a Gaussian's exponent).  Let us define two Gaussian basis elements, $\ket{g_i}$ and $\ket{g_j}$, by their real-space representations,
\begin{equation}
g_i(\vc{r}) = N e^{-(\vc{r} - \vc{r}_A)\mx{\alpha}(\vc{r} - \vc{r}_A)} \quad , \quad
g_j(\vc{r}) = N' e^{-(\vc{r} - \vc{r}_B)\mx{\beta}(\vc{r} - \vc{r}_B)} \label{eqDefineGaussians}
\end{equation}
Vectors $\vc{r}_A$ and $\vc{r}_B$ are the center positions of $\ket{g_i}$ and $\ket{g_j}$, and $\mx{\alpha}$ and $\mx{\beta}$ are $d\times d$ diagonal matrices specifying their exponential factors.  We will make repeated use of the identity
\begin{equation}
g_i(\vc{r}) g_j(\vc{r}) = N N' K e^{ -(\vc{r} - \vc{R}_{AB})\mx{\mu}(\vc{r} - \vc{R}_{AB}) } \label{eqCombineGaussian}
\end{equation}
where
\begin{eqnarray}
K &=& e^{-(\vc{r}_A-\vc{r}_B)\mx{C}(\vc{r}_A-\vc{r}_B)} \quad \mbox{with}\quad \mx{C} = \mx{\alpha}\mx{\beta}/(\mx{\alpha}+\mx{\beta}) \\ \label{eqK}
\vc{R}_{AB} &=& (\mx{\alpha} \vc{r}_A + \mx{\beta} \vc{r}_B)/(\mx{\alpha} + \mx{\beta}) \\ \label{eqRAB}
\mx{\mu} &=& \mx{\alpha} + \mx{\beta} \label{eqMu}
\end{eqnarray}
which allows us to transform the product of two Gaussians into a single Gaussian.  Throughout this appendix, division by a matrix means multiplication by its inverse.

%  As a final preliminary, we define for convenience the matrix $\mx{P}\equiv \mx{\alpha} \mx{\beta}$.
Next, let us define a ``piecewise-polynomial'' operator $\mathcal{O}$ as one which can be written in the form
\begin{eqnarray}
\mathcal{O} &=& \sum_i O_i(\vc{r}) \chi_{[\vc{r}_i,\vc{r}_{i+1}]}(\vc{r}) \label{eqPiecewisePoly}\\
&=& \sum_i \left( \sum_t^{N_i} c_t \prod_k^d r_k^{n(k,t)} \right) \chi_{[\vc{r}_i,\vc{r}_{i+1}]}(\vc{r})\,.
\end{eqnarray}
$O_i$ is polynomial in the components of $\vc{r}$, and is expanded in polynomial terms in the second line.  The characteristic function $\chi_{[\vc{a},\vc{b}]}$ equals 1 within the $d$-dimensional interval $[\vc{a},\vc{b}]$ (\emph{e.g.} a cube for $d=3$) and 0 everywhere else.  The components of $\vc{a}$ and $\vc{b}$ are real or $\pm\infty$.  Thus, $\langle g_i | \mathcal{O} | g_j \rangle$ is a linear combination of terms with the form $\langle g_i| \chi_{[\vc{a},\vc{b}]}(\vc{r}) \prod_{k=1}^d r_k^{n_k} | g_j \rangle$ where $\vc{n}$, $\vc{a}$, and $\vc{b}$ are $d$-dimensional vectors.  We compute this general element immediately and then use the result to find the matrix elements of the kinetic and potential energy operators used for DQD Hamiltonians.

We begin the computation as follows,
\begin{eqnarray}
\bra{g_i} \chi_{[\vc{a},\vc{b}]}(\vc{r}) \prod_{k=1}^d r_k^{n_k}  \ket{g_j} 
&=& NN'K \int_{\vc{a}'}^{\vc{b}'} d^d\vc{u} e^{-\vc{u}\mx{\mu} \vc{u}} \prod_{k=1}^d \left(u_k + R_{ABk}\right)^{n_k} \label{eqPolyDeriv1} \\
&=&  NN'K \prod_{k=1}^d \int_{a_k'}^{b_k'} du_k e^{-\mu_{kk}u_k^2} \left(u_k + R_{ABk}\right)^{n_k} \\
&=& NN'K \prod_{k=1}^d \sum_{l=0}^{n_k} {n_k \choose{l}}\left(R_{ABk}\right)^{n_k-l} \int_{a_k'}^{b_k'} du_k u_k^l e^{-\mu_{kk}u_k^2} \label{eqPolyDeriv2} \\
&=& NN'K \prod_{k=1}^d \sum_{l=0}^{n_k} {n_k \choose{l}}\left(R_{ABk}\right)^{n_k-l} F(l;a_k',b_k',\mu_{kk})  \label{eqPolyDeriv3}
\end{eqnarray}
In the first line we have used \eqref{eqCombineGaussian} and made the substitution $\vc{u}=\vc{r}-\vc{R}_{AB}$.  The limits of integration are from $\vc{a}' = \vc{a} - \vc{R}_{AB}$ to $\vc{b}' = \vc{b} - \vc{R}_{AB}$.  In the second line the integral written as product of one-dimensional integrals, and in the third line the binomial expansion of $(u_i+R_{ABi})^{n_i}$ is inserted.  The fourth line (Eq. \ref{eqPolyDeriv3}) defines the integral
\begin{equation}
F(l;a,b,\mu) \equiv \int_a^b du u^l e^{-\mu u^2}
\end{equation}
which we now compute.

Using the shorthand notation $F(l)$ for $F(l;a,b,\mu)$,and integrating by parts ($U=u^{l-1}$, $dV=ue^{\mu u^2}du$, $dU=(l-1)u^{l-2} du$, $V=-1/(2\mu)e^{-\mu u^2}$)
\begin{eqnarray}
F(l) &=& \left. \frac{-u^{l-1}}{2\mu}e^{-\mu u^2} \right]_a^b + \frac{l-1}{2\mu}\int_a^b u^{l-2}e^{-\mu u^2} du \\
&=& G(l) + \frac{l-1}{2\mu}F(l-2) \quad \mbox{where} \,\, G(l) \equiv \left. \frac{-u^{l-1}}{2\mu}e^{-\mu u^2} \right]_{u=a}^b
\end{eqnarray}
Since 
\begin{eqnarray}
F(0) &=& \int_a^b e^{-\mu u^2} du = \frac{1}{2}\sqrt{\frac{\pi}{\mu}}\left[ \mbox{erf}(\sqrt{\mu}b) - \mbox{erf}(\sqrt{\mu}a) \right] \quad \mbox{and}\\
F(1) &=& \int_a^b ue^{-\mu u^2} du = \frac{1}{2\mu}\left[ e^{-\mu a^2} - e^{-\mu b^2} \right]
\end{eqnarray}
can be computed directly, we can write $F(l)$ non-recursively by
\begin{equation}
F(l) = \sum_{m=0}^{m_{max}} G(l-2m) \prod_{p=1}^m \frac{l-2p+1}{2\mu} + F^*\prod_{p=1}^{l/2}\frac{l-2p+1}{2\mu} \quad (l \mbox{ even})
\end{equation}
where 
\begin{equation}
m_{max} = \left\{ 
\begin{array}{cl}
l/2-1 & l\mbox{ even} \\
(l-1)/2-1 & l\mbox{ odd} 
\end{array}
\right.
\quad \mbox{and} \quad
 F^* = \left\{ 
\begin{array}{cl}
F(0) & l\mbox{ even} \\
F(1) & l\mbox{ odd} 
\end{array}
\right. \,.
\end{equation}

In practice, $F(l)$ is calculated by a routine which iteratively builds the solution, such as the following (pseudo C++):

\begin{center}
\begin{verbatim}
real F(k) {
  i = ( k is even ) ? 0 : 1
  x=F(i)

  while( i < k ) {
     i++
     x *= i/(2*mu)
     i++
     x += G(i)
  }
  return x
}
\end{verbatim}
\end{center}
With the above analytic expression of $F(l;a,b,\mu)$, Eq.~\ref{eqPolyDeriv3} can be used to compute $\langle g_i| \chi_{[\vc{a},\vc{b}]}(\vc{r}) \prod_{k=1}^d r_k^{n_k} | g_j \rangle$ and thus the matrix elements of any ``piecewise-polynomial'' operator $\mathcal{O}$.

%special case when pure polynomical -- remove??
Note that if the matrix element is a single polynomial, so that $a_k = -\infty$ and $b_k = +\infty$ for $k = 1\ldots d$, then $F(l)$ is zero for odd $l$ (integrand is odd) and for even $l$ is given by the elementary integral 
\begin{equation}
\int_{-\infty}^\infty x^{2n} e^{-ax^2} dx = 2\sqrt{\pi}\,\frac{2n!}{n!}\left(\frac{1}{2\sqrt{a}}\right)^{2n+1}
\end{equation}
Thus, in this simplified case when there are no characteristic functions,
\begin{equation}
\bra{g_i} \prod_{k=1}^d r_k^{n_k}  \ket{g_j} =
NN'K \prod_{k=1}^d \left[ \sum_{l=0, \mathrm{even}}^{n_k} {n_k \choose{l}}\left(R_{ABk}\right)^{n_k-l}\left( 2\sqrt{\pi} \frac{l!}{(l/2)!}\left(\frac{1}{2\sqrt{\mu_{kk}}}\right)^{l+1}\right)\right] \label{eqPurePolyCase}
\end{equation}

Piecewise-polynomial potentials, such as parabolic or quartic quantum dot potentials, are naturally expressed in the form given by Eq.~\ref{eqPiecewisePoly}.  Hamiltonian terms involving the momentum operator, however, require a few preliminary steps.  For example, matrix elements of the kinetic energy operator, $\bra{g_i} -\nabla^2 \ket{g_j}$, can written as matrix elements of the operator $(\vc{r}-\vc{r}_A) (4\mx{\alpha}\mx{\beta}) (\vc{r} - \vc{r}_B)$, which is polynomial in $\vc{r}$ by integrating by parts and taking derivatives of the Gaussian basis functions.
\begin{eqnarray}
\bra{g_i} -\nabla^2 \ket{g_j} 
&=&  \int d^d\vc{r}\, \nabla \overline{g_i(\vc{r})} \cdot \nabla g_j(\vc{r}) \label{eqKineticDeriv1} \\
&=&  \int d^d\vc{r}\, \overline{g_i(\vc{r})} g_j(\vc{r}) (\vc{r}-\vc{r}_A) (4\mx{\alpha}\mx{\beta}) (\vc{r} - \vc{r}_B) \label{eqKineticDeriv2} \\
&=& \bra{g_i} (\vc{r}-\vc{r}_A) (4\mx{\alpha}\mx{\beta}) (\vc{r} - \vc{r}_B) \ket{g_j}
\end{eqnarray}
Next define diagonal matrix $\mx{P} \equiv \mx{\alpha}\mx{\beta}$, and use Eq.~\ref{eqPurePolyCase} to arrive at a formula in terms of the basis element parameters
\begin{eqnarray}
\bra{g_i} -\nabla^2 \ket{g_j} 
&=& 4 \bra{g_i} \vc{r}\mx{P}\vc{r} - \vc{r}\mx{P}\vc{r}_B - \vc{r}_A\mx{P}\vc{r} + \vc{r}_A\mx{P}\vc{r}_B \ket{g_j} \\
&=& 4NN'K \sum_k^d \left[ P_{kk}\left( R_{ABk}^2\sqrt{\frac{\pi}{\mu_{kk}}} + \frac{\sqrt{\pi}}{2\mu_{kk}^{3/2}} \right)\prod_{l\ne k} \sqrt{\frac{\pi}{\mu_{ll}}} \right. \\
&& \left. + \,\left(r_{Ak}P_{kk}r_{Bk} - P_{kk}r_{Bk} - r_{Ak}P_{kk}\right)\prod_{l} \sqrt{\frac{\pi}{\mu_{ll}}} \right] \\
&=& \mathcal{K} \sum_k^d P_{kk}\left[ \frac{\sqrt{\pi}}{2\mu_{kk}^{3/2}}\prod_{l\ne k} \sqrt{\frac{\pi}{\mu_{ll}}} +  \left(R_{ABk}^2 -(r_{Ak}+r_{Bk}) + r_{Ak}r_{Bk}\right)\prod_{l} \sqrt{\frac{\pi}{\mu_{ll}}} \right] \label{eqFirstKappa}\\
&=& \mathcal{K} \sum_{k=1}^d P_{kk} \left(\frac{\sqrt{\pi}}{2\mu_{kk}^{3/2}}\right)\prod_{l\ne k} \sqrt{\frac{\pi}{\mu_{ll}}} +  \vc{\Delta}_A\mx{P}\vc{\Delta}_B\prod_{k=1}^d \sqrt{\frac{\pi}{\mu_{kk}}}
\end{eqnarray}
We define $\mathcal{K}=4NN'K$ in Eq.~\ref{eqFirstKappa}, and in the last line we have defined $\vc{\Delta}_A \equiv \vc{R}_{AB}-\vc{r}_A$ and $\vc{\Delta}_B \equiv \vc{R}_{AB}-\vc{r}_B$.

%Alternate (original) derivation of kinetic element -- remove???
The kinetic matrix element may also be computed directly as follows:

\medskip
\lefteqn{\bra{g_i} -\nabla^2 \ket{g_j}}
\vspace{-0.5cm}
\begin{eqnarray}
&=&  \int d^d\vc{r}\, \nabla g_i(\vc{r}) \cdot \nabla g_j(\vc{r}) \label{eqAltKineticDeriv1} \\
&=& 4NN' \int d^d\vc{r} e^{-(\vc{r} - \vc{r}_A)\mx{\alpha}(\vc{r} - \vc{r}_A)} e^{-(\vc{r} - \vc{r}_B)\mx{\alpha}(\vc{r} - \vc{r}_B)} (\vc{r}-\vc{r}_A) \mx{P} (\vc{r} - \vc{r}_B) \label{eqAltKineticDeriv2} \\
&=& \mathcal{K} \int d^d\vc{r} e^{-(\vc{r} - \vc{R}_{AB})\mx{\mu}(\vc{r} - \vc{R}_{AB})} \left[(\vc{r}- \vc{R}_{AB})+(\vc{R}_{AB}- \vc{r}_A)\right] \mx{P} \left[(\vc{r} - \vc{R}_{AB}) + (\vc{R}_{AB}-\vc{r}_B)\right] \label{eqKineticDeriv3} \\
&=& \mathcal{K} \int d^d\vc{u} e^{-\vc{u}\mx{\mu} \vc{u}} \left[\vc{u}\mx{P}\vc{u} + \vc{u}\mx{P}\frac{\mx{\alpha}-\mx{\beta}}{\mx{\alpha}+\mx{\beta}}(\vc{r}_A-\vc{r}_B) + (\vc{R}_{AB}- \vc{r}_A) \mx{P} (\vc{R}_{AB}-\vc{r}_B) \right] \label{eqKineticDeriv4} \\
&=& \mathcal{K} \left[ \int d^d\vc{u} e^{-\vc{u}\mx{\mu} \vc{u}} \vc{u}\mx{P}\vc{u} + \vc{\Delta}_A \mx{P} \vc{\Delta}_B \int d^d\vc{u} e^{-\vc{u}\mx{\mu} \vc{u}} \right] \label{eqKineticDeriv5} \\
&=& \mathcal{K} \left[ \sum_{i=1}^d \int d^d\vc{u} P_{ii}u_i^2 e^{-\vc{u}\mx{\mu} \vc{u}}  + \vc{\Delta}_A \mx{P} \vc{\Delta}_B \int d^d\vc{u} e^{-\vc{u}\mx{\mu} \vc{u}} \right] \label{eqKineticDeriv6} \\
&=& \mathcal{K} \left[ \sum_{i=1}^d P_{ii} \left( \int du_i u_i^2 e^{-\mu_{ii}u_i^2} \right)\left(\prod_{j\ne i}\int du_j e^{-\mu_{jj}u_j^2}\right)  + \vc{\Delta}_A \mx{P} \vc{\Delta}_B \prod_{i=1}^d \int du_i e^{-\mu_{ii}u_i^2} \right] \label{eqKineticDeriv7} \\
&=& \mathcal{K} \left[ \sum_{i=1}^d P_{ii}\left(\frac{\sqrt{\pi}}{2\mu_{ii}^{3/2}}\right)\prod_{j\ne i} \sqrt{\frac{\pi}{\mu_{jj}}} + \vc{\Delta}_A \mx{P} \vc{\Delta}_B \prod_{i=1}^d \sqrt{\frac{\pi}{\mu_{ii}}} \right] \label{eqKineticDeriv8}
\end{eqnarray}
We integrate by parts to get \eqref{eqAltKineticDeriv1}, use \eqref{eqCombineGaussian} to transition to \eqref{eqKineticDeriv3}, and substitute $\vc{u}=\vc{r}-\vc{R}_{AB}$ in \eqref{eqKineticDeriv4}.  The term of \eqref{eqKineticDeriv4} linear in $u$ vanishes since the integrand is odd.  We then expand the matrix notation to arrive at \eqref{eqKineticDeriv7}.  The final line is obtained using the elementary integrals
\begin{eqnarray}
\int_{-\infty}^\infty e^{ax^2} dx &=& \sqrt{\frac{\pi}{a}} \\
\int_{-\infty}^\infty x^2 e^{ax^2} dx &=& \frac{\sqrt{\pi}}{2a^{3/2}}
\end{eqnarray}

The overlap matrix element can be calculated using Eq.~\ref{eqPolyDeriv3} or \ref{eqPurePolyCase}, though it is straightforward to compute directly by using Eq.~\ref{eqCombineGaussian} and the translation invariance of the integral.  Following the latter approach, we obtain
\begin{equation}
\bra{g_i} g_j \rangle = NN'\int d^d\vc{r}\, K e^{-\vc{r}\mx{\mu} \vc{r}} = NN'K\sqrt{\frac{\pi^d}{\det{\mx{\mu}}}}
\end{equation}

\subsection{Coulomb matrix elements \label{secCoulombMatrixEls}}
To compute the matrix elements of the $m$-body Hamiltonian \eqref{eqMBHam} in the basis $\Bmb_m$, matrix elements for the two-particle Coulomb term $\frac{e^2}{\kappa r}$ must be computed.  We introduce two more Gaussian basis elements, $\ket{g_{i'}}$ and $\ket{g_{j'}}$ written similar to those in \eqref{eqDefineGaussians}:
\begin{equation}
g_{i'}(\vc{r}) = N'' e^{-(\vc{r} - \vc{r}_C)\mx{\gamma}(\vc{r} - \vc{r}_C)} \quad , \quad
g_{j'}(\vc{r}) = N''' e^{-(\vc{r} - \vc{r}_D)\mx{\delta}(\vc{r} - \vc{r}_D)}
\end{equation}
Anticipating the combination of $g_{i'}(\vc{r})$ and $g_{j'}(\vc{r})$ using \eqref{eqCombineGaussian}, define the following analogous to Eqs.~\eqref{eqK}-\eqref{eqMu}:
\begin{eqnarray}
K' &=& e^{-(\vc{r}_C-\vc{r}_D)\mx{C}(\vc{r}_C-\vc{r}_D)} \quad \mbox{with}\quad C = \mx{\gamma}\mx{\delta}/(\mx{\gamma}+\mx{\delta}) \\ 
\vc{R}_{CD} &=& (\mx{\gamma} \vc{r}_C + \mx{\delta} \vc{r}_D)/(\mx{\gamma} + \mx{\delta}) \\
\mx{\nu} &=& \mx{\gamma} + \mx{\delta}
\end{eqnarray}

\noindent We now turn to the matrix element of interest. We begin by writing it as a real-space integral then apply Eq.~\ref{eqCombineGaussian} to each pair of Gaussian basis elements:
\begin{eqnarray}
\bra{g_i g_{i'}} \frac{e^2}{\kappa r} \ket{g_j g_{j'}} &=& \frac{e^2}{\kappa} \int d^d\vc{r}_1 d^d\vc{r}_2\, g_i(\vc{r}_1)g_{i'}(\vc{r}_2) \frac{1}{r_{12}} g_j(\vc{r}_1)g_{j'}(\vc{r}_2) \\
&=& \mathcal{K} \frac{e^2}{\kappa} \int d^d\vc{r}_1 d^d\vc{r}_2\, e^{-(\vc{r}_1 - \vc{R}_{AB})\mx{\mu}(\vc{r}_1 - \vc{R}_{AB})} \frac{1}{r_{12}} e^{-(\vc{r}_2 - \vc{R}_{AB})\mx{\mu}(\vc{r}_2 - \vc{R}_{AB})}  \label{eqCoulMxEl1}
\end{eqnarray}
where $\mathcal{K} \equiv NN'N''N'''KK'$.  Next, we write each of the two exponentials and $1/r_{12}$ in terms of their Fourier transforms, defined by $\mathrm{FT}\left[ f(\vc{r}) \right] = (2\pi)^{-d} \int d^d\vec{k} f(\vc{k}) e^{i\vc{k}\cdot\vc{r}}$.  At this point we fix $d=3$, so that the Fourier transform of $1/r$ is well defined.  The 2D solution will be obtained later, by taking a limit of the 3D result. Thus, the Fourier transforms
\begin{eqnarray}
\mathrm{FT}\left[ e^{-\vc{x}\mx{\alpha} \vc{x}} \right] &=& \left(\sqrt{\frac{\pi^d}{\det \mx{\alpha}}}\right)e^{-\frac{1}{4}\vc{k} \mx{\alpha}^{-1} \vc{k}} \quad \mbox{and}\\
\mathrm{FT}\left[ \frac{1}{r} \right] &=& \frac{4\pi}{k^2} \qquad \mbox{for $d=3$}
\end{eqnarray}
are inserted into in Eq.~\ref{eqCoulMxEl1} and result in (for $d=3$)
\begin{eqnarray}
\bra{g_i g_{i'}} \frac{e^2}{\kappa r} \ket{g_j g_{j'}} &=& \frac{\mathcal{K}e^2}{(2\pi)^{3d}\kappa} \int d^d\vc{r}_1 d^d\vc{r}_2 d^d\vc{k}_1 d^d\vc{k}_2 d^d\vc{k}_3 \sqrt{\frac{\pi^d}{\det \mx{\mu}}} e^{-\frac{1}{4}\vc{k}_1 \mx{\mu}^{-1} \vc{k}_1} \\ 
& & \hspace{-1cm}\times\, \frac{4\pi}{k_2^2} \sqrt{\frac{\pi^d}{\det \mx{\nu}}}e^{-\frac{1}{4}\vc{k}_3 \mx{\nu}^{-1} \vc{k}_3} e^{i\vc{k}_1\cdot(\vc{r}_1-\vc{R}_{AB})} e^{i\vc{k}_2\cdot(\vc{r}_1-\vc{r}_2)}e^{i\vc{k}_3\cdot(\vc{r}_2-\vc{R}_{CD})} \label{eqCoulMxEl2}
\end{eqnarray}
Integrating over $\vc{r}_1$ and $\vc{r}_2$ yields delta functions $(2\pi)^d\delta(\vc{k}_1+\vc{k}_2)$ and $(2\pi)^d\delta(\vc{k}_2-\vc{k}_3)$.  Then integrating over $\vc{k}_2$ and $\vc{k}_3$ effectively sets $-\vc{k}_1=\vc{k}_2=\vc{k}_3$ in the integrand, and we define $\vc{k}\equiv -\vc{k}_1$ to clean up the notation. After these integrations, Eq.~\ref{eqCoulMxEl2} becomes
\begin{equation}
\bra{g_i g_{i'}} \frac{e^2}{\kappa r} \ket{g_j g_{j'}} = \frac{\mathcal{K}e^2}{(2\pi)^d\kappa} \frac{\pi^d}{\sqrt{\det \mx{\mu}\mx{\nu}}} \int d^d\vc{k} \frac{4\pi}{k^2} e^{-\vc{k} \mx{\sigma} \vc{k}} e^{i\vc{k}\cdot\vc{\Delta}} \label{eqCoulMxEl3}
\end{equation}
where $\mx{\sigma} \equiv \frac{\mx{\mu} + \mx{\nu}}{4\mx{\mu}\mx{\nu}}$ (division is multiplication by matrix inverse) and $\vc{\Delta} = \vc{R}_{AB} - \vc{R}_{CD}$ .  Define $\trsigma = \mathrm{Tr}(\mx{\sigma})/d$, write $e^{-\vc{k}\mx{\sigma} \vc{k}} = \exp(-\trsigma k^2)\exp\left[-\vc{k}(\mx{\sigma}-\trsigma\mathbf{1})\vc{k}\right]$ ($\mathbf{1}$ is the identity matrix in $d$ dimensions), and use the identity
\begin{equation}
\frac{e^{-\trsigma k^2}}{k^2} = 2\trsigma \int_0^1 dS S^{-3} e^{-\trsigma k^2/S^2}
\end{equation}
(this follows from $e^{-a} = a \int_1^\infty e^{-ax}dx$ with $x=1/S^2$ and $a=\trsigma k^2$) to transform Eq.~\ref{eqCoulMxEl3} into
  \begin{eqnarray}
\bra{g_i g_{i'}} \frac{e^2}{\kappa r} \ket{g_j g_{j'}} &=& \frac{\mathcal{K}e^2}{\kappa} \frac{4\pi}{(2\pi)^d}\frac{\pi^d}{\sqrt{\det \mx{\mu}\mx{\nu}}} 2\trsigma \int_0^1 \frac{dS}{S^3} \int d^d\vc{k} e^{-\vc{k} (\mx{\sigma} - \trsigma + \trsigma/S^2) \vc{k}} e^{i\vc{k}\cdot\vc{\Delta}} \\
&=& \frac{\mathcal{K}e^2}{\kappa} \frac{4\pi}{(2\pi)^d}\frac{\pi^d}{\sqrt{\det \mx{\mu}\mx{\nu}}} 2\trsigma \int_0^1 \frac{dS}{S^3} \frac{\pi^{d/2} e^{-\frac{1}{4}\vc{\Delta} (\mx{\sigma} - \trsigma + \trsigma/S^2)^{-1} \vc{\Delta}}}{\sqrt{\det (\mx{\sigma} - \trsigma + \trsigma/S^2)}}  \\
&=& \frac{\mathcal{K}e^2}{\kappa} \frac{\pi^{d/2+1}\trsigma}{2^{d-3}}\left(\det \mx{\mu}\mx{\nu}\right)^{-\frac{1}{2}} I(\mx{\sigma},\vc{\Delta}) \label{eqCoulMxElFirstI}\\ 
&=& \frac{\mathcal{K}e^2}{\kappa} \pi^{5/2}\trsigma \left(\det \mx{\mu}\mx{\nu}\right)^{-\frac{1}{2}} I(\mx{\sigma},\vc{\Delta}) \label{eqCoulMxEl4}
\end{eqnarray}
where $\trsigma \equiv \trsigma\mathbf{1}$ when used in a matrix context, and in the last line we explicitly put $d=3$.  In Eq.~\ref{eqCoulMxElFirstI} we have defined the integral
\begin{eqnarray}
I(\mx{\sigma},\vc{\Delta}) &=& \int_0^1 \frac{dS}{S^3} \left[ \det (\mx{\sigma} - \trsigma + \trsigma/S^2)\right]^{-\frac{1}{2}} e^{-\frac{1}{4}\vc{\Delta} (\mx{\sigma} - \trsigma + \trsigma/S^2)^{-1} \vc{\Delta}} \label{eqI1}\\
&=& \int_0^1 \frac{dS}{S^3} \left[ \prod_{i=1}^3 (\sigma_{ii} - \trsigma(1 - 1/S^2))\right]^{-\frac{1}{2}} e^{-\frac{1}{4}\sum_{i=1}^3 \vc{\Delta}_i^2/(\sigma_{ii} - \trsigma(1 - 1/S^2))} \label{eqI2}
\end{eqnarray}
The second line follows since $\mx{\sigma}$ is a $3\times3$ diagonal matrix. We cannot express the integral in closed form, and so must compute $I(\mx{\sigma},\vc{\Delta})$ numerically.  In the two-dimensional case, we take the limit $\mu_{33} = \nu_{33} \rightarrow \infty$, which means that $\sigma_{33} \rightarrow 0$.  This is the limit where all the Gaussian basis elements have the same width $\eta$ in the $z$-direction, and $\eta$ approaches zero (making the elements two-dimensional).  $\mathcal{K}$ contains the factor $(2\alpha_{33}/\pi)^{1/4}(2\beta_{33}/\pi)^{1/4}(2\gamma_{33}/\pi)^{1/4}(2\delta_{33}/\pi)^{1/4} = 2\eta/\pi$, which cancels the factor $1/\sqrt{\mu_{33}\nu_{33}} = 1/\sqrt{(2\eta)(2\eta)} = 1/(2\eta)$ from $(\det \mx{\mu}\mx{\nu})^{-1/2}$, leaving a factor of $\pi$ in the denominator.  This reduces the exponent of $\pi$ in Eq.~\ref{eqCoulMxEl4} from $5/2$ to $3/2$ as seen below in Eq.~\ref{eqFinalCoulomb2D}.

The integral $I(\mx{\sigma},\vc{\Delta})$ converges.  The only possible trouble occurs when $(\sigma_{ii} - \trsigma(1 - 1/S^2)) = 0$, or equivalently $S=1/\sqrt{1-\sigma_{ii}/\trsigma}$, for some $i=1,2,3$.  Since $\sigma_{ii} > 0$ implies that $1/\sqrt{1-\sigma_{ii}/\trsigma}$ is either imaginary or greater than 1, the only divergence of the integrand for $S\in [0,1]$ occurs at $S=1$ when $\sigma_{ii} = 0$.  Although this happens for $i=3$ in the 2D case ($\sigma_{33} = 0$), the divergence is $\sim 1/\sqrt{1-S}$, which is integrable.  Thus, the integral is well defined and convergent over the entire range of physical parameters.

In summary, the Coulomb term matrix elements for two and three dimensions can be carried out analytically up to the numerical evaluation of a convergent one-dimensional integral.  They are given by:
\begin{eqnarray}
\bra{g_i g_{i'}} \frac{e^2}{\kappa r} \ket{g_j g_{j'}} &=& \frac{\mathcal{K}e^2}{\kappa} \pi^{5/2}\trsigma\left(\det \mx{\mu}\mx{\nu}\right)^{-\frac{1}{2}} I(\mx{\sigma},\vc{\Delta}) \quad \mbox{(3D)} \label{eqFinalCoulomb3D}\\
\bra{g_i g_{i'}} \frac{e^2}{\kappa r} \ket{g_j g_{j'}} &=& \frac{\mathcal{K}_{2D}e^2}{\kappa} \pi^{3/2}\trsigma\left(\det_{2D} \mx{\mu}\mx{\nu}\right)^{-\frac{1}{2}} I(\mx{\sigma},\vc{\Delta}) \quad \mbox{(2D)} \label{eqFinalCoulomb2D}
\end{eqnarray}
The subscripts ``2D'' are a reminder that the normalization factors and determinant contain only $x$ and $y$ factors.  

%Additional implementation info -- remove??
\medskip
\noindent\textbf{Implementation Note:}\\
We note additionally that when the integral for two dimensions is computed numerically, the integral around $S=1$ is approximated by a closed form.  At $S=1-\epsilon$, when $\epsilon \ll 1$, the integrand in Eq.~\ref{eqI1}, which we denote $W(S)$, is approximated via the expansion
\begin{equation}
\mx{\sigma} - \trsigma\left(1-\frac{1}{(1-\epsilon)^2}\right) \approx \mx{\sigma} + 2\trsigma\epsilon
\end{equation}
resulting in an approximation for the integral between $S=1-\epsilon_0$ and $1$.
\begin{eqnarray}
\int_{1-\epsilon}^1 dS\, W(S) & \approx & \int_{1-\epsilon}^1 dS \det\left(\mx{\sigma} + \trsigma(2\epsilon)\right)^{-1/2} (1-\epsilon)^{-3} e^{-\frac{1}{4}(\Delta_x^2\sigma_x^{-1} + \Delta_y^2\sigma_y^{-1})} \\
&=& \int_0^{\epsilon_0} d\epsilon \left(\sigma_x\sigma_y\trsigma(2\epsilon)\right)^{-1/2} e^{-\frac{1}{4}(\Delta_x^2\sigma_x^{-1} + \Delta_y^2\sigma_y^{-1})} \\
&=& \frac{e^{-\frac{1}{4}(\Delta_x^2\sigma_x^{-1} + \Delta_y^2\sigma_y^{-1})}}{\sqrt{2\sigma_x\sigma_y\trsigma}} \int_0^{\epsilon_0} \frac{d\epsilon}{\sqrt{\epsilon}} \\
&=& \frac{\sqrt{2\epsilon_0} e^{-\frac{1}{4}(\Delta_x^2\sigma_x^{-1} + \Delta_y^2\sigma_y^{-1})}}{\sqrt{\sigma_x\sigma_y\trsigma}}
\end{eqnarray}

\subsection{Isotropic Gaussian functions - the completely analytic case}
When all the Gaussian basis elements are isotropic, that is, when $\mx{\alpha}$ is a proportional to the identity, then Coulomb matrix elements can be computed analytically in both two and three dimensions.  In 3D, we follow the above derivation up to Eq.~\ref{eqCoulMxEl3} and let $\mx{\mu} = \mu\mathbf{1}$, $\mx{\nu} = \nu\mathbf{1}$, and $\mx{\sigma} = \sigma\mathbf{1}$ define the scalars $\mu$, $\nu$, and $\sigma$ corresponding to the similarly named matrix.  Then write Eq.~\ref{eqCoulMxEl3} in spherical coordinates with the $z$-axis along $\vc{\Delta}$ to obtain
\begin{eqnarray}
\bra{g_i g_{i'}} \frac{e^2}{\kappa r} \ket{g_j g_{j'}} 
&=& \mathcal{K}' \int k^2 dk\, d(\cos\theta)\, d\phi\, \frac{4\pi}{k^2} e^{-\sigma k^2} e^{ik\Delta \cos\theta} \\
&=& 4\pi\mathcal{K}' \int_0^\infty dk\,e^{-\sigma k^2} \int_{-1}^1 d(\cos\theta)\, e^{ik\Delta \cos\theta} \int_0^{2\pi}d\phi  \\
&=& 8\pi\mathcal{K}' \int_0^\infty dk\, \frac{e^{-\sigma k^2}}{ik\Delta} \left(e^{ik\Delta}-e^{-ik\Delta}\right) \\
&=& 16\pi\mathcal{K}'\int_0^\infty dk\, e^{-\sigma k^2} \frac{\sin k\Delta}{k} \\
&=& 16\pi\mathcal{K}' \int_0^\Delta dy\,\int_0^\infty dk\, e^{-\sigma k^2}\cos ky \label{eqFirstY}\\
&=& 8\pi\mathcal{K}' \int_0^\Delta dy\,\int_{-\infty}^\infty dk\, e^{-\sigma k^2}\cos ky
\end{eqnarray}
where $\mathcal{K}' = \frac{\mathcal{K}e^2}{8\kappa} (\mu\nu)^{-3/2}$.  In line \ref{eqFirstY} we introduce a dummy variable $y$ in order to make the $k$-integral tractable (see below).  The final line uses the fact that the integrand is even to extend the range of integration.  By writing $\cos ky$ in exponential form and completing the squares, we can integrate over $k$ to get
\begin{eqnarray}
\bra{g_i g_{i'}} \frac{e^2}{\kappa r} \ket{g_j g_{j'}} 
&=& 4\pi\mathcal{K}' \int_0^\Delta dy\,\int_{-\infty}^\infty dk\, e^{-\sigma k^2}\left( e^{iky} + e^{-iky} \right) \\
&=& 4\pi\mathcal{K}' \int_0^\Delta dy\,\int_{-\infty}^\infty dk\, \left( e^{-\sigma\left(k-\frac{iy}{2\sigma}\right)^2} + e^{-\sigma\left(k+\frac{iy}{2\sigma}\right)^2} \right) e^{-\frac{y^2}{4\sigma}} \\
&=& 4\pi\mathcal{K}' \int_0^\Delta dy\,\int_{-\infty}^\infty dk\, 2e^{-\sigma k^2} e^{-\frac{y^2}{4\sigma}} \label{eqCoulMxElShift}\\
&=& 8\pi\mathcal{K}'\sqrt{\frac{\pi}{\sigma}} \int_0^\Delta dy\, e^{-\frac{y^2}{4\sigma}} \\
&=& 8\pi\mathcal{K}'\sqrt{\frac{\pi}{\sigma}} \left[ \frac{\sqrt{\pi/(4\sigma)}}{2}\mbox{erf}(\Delta) \right] \label{eqCoulMxElErf} \\
&=& \mathcal{K}'\frac{2\pi^2}{\sigma}\mbox{erf}(\Delta)
\end{eqnarray}
We obtain line \ref{eqCoulMxElShift} by shifting $k$ in each of the terms (in different directions), which does not alter the integral and introduce the error function $\mbox{erf}(x) = \frac{2}{\sqrt{\pi}}\int_0^x dx'\,e^{-x'^2}$ on line \ref{eqCoulMxElErf}.  Expanding $\mathcal{K}'$ gives us a final analytic formula for Coulomb matrix elements in three dimensions ($d=3$),
\begin{equation}
\bra{g_i g_{i'}} \frac{e^2}{\kappa r} \ket{g_j g_{j'}} = \frac{\mathcal{K}e^2}{4\kappa\sigma} \frac{\pi^2}{(\mu\nu)^{3/2}}\,\mbox{erf}(\Delta) \qquad \mbox{(3D, isotropic)}
\end{equation}

In 2D, we proceed to Eq.~\ref{eqCoulMxEl1} as above, but then insert 2D Fourier transforms instead of the 3D ones yielding Eq.~\ref{eqCoulMxEl2}.  The Fourier transform of the Gaussian basis elements has the same form as the 3D case, but now 
\begin{equation}
\mathrm{FT}\left[ \frac{1}{r} \right] = \frac{2\pi}{k} \qquad \mbox{for $d=2$} 
\end{equation}
Inserting these into to Eq.~\ref{eqCoulMxEl1} (with $d=2$) gives
\begin{eqnarray}\bra{g_i g_{i'}} \frac{e^2}{\kappa r} \ket{g_j g_{j'}} &=& \frac{\mathcal{K}e^2}{(2\pi)^{6}\kappa} \int d^2\vc{r}_1 d^2\vc{r}_2 d^2\vc{k}_1 d^2\vc{k}_2 d^2\vc{k}_3 \sqrt{\frac{\pi^2}{\mu^2}} e^{-k_1^2/(4\mu)} \\ 
& & \hspace{-1cm}\times\, \frac{2\pi}{k_2} \sqrt{\frac{\pi^2}{\nu^2}}e^{-k_3^2/(4\nu)} e^{i\vc{k}_1\cdot(\vc{r}_1-\vc{R}_{AB})} e^{i\vc{k}_2\cdot(\vc{r}_1-\vc{r}_2)}e^{i\vc{k}_3\cdot(\vc{r}_2-\vc{R}_{CD})}
\end{eqnarray}
As in the 3D case, integrating over $\vc{r}_1$ and $\vc{r}_2$ produces delta functions which are removed by then integrating over $\vc{k}_2$ and $\vc{k}_3$.  This effectively sets $\vc{k}\equiv-\vc{k}_1=\vc{k}_2=\vc{k}_3$ in the integrand,
\begin{equation}
\bra{g_i g_{i'}} \frac{e^2}{\kappa r} \ket{g_j g_{j'}} = \frac{\mathcal{K}e^2}{4\kappa} \frac{1}{\mu\nu} \int d^2\vc{k}\, \frac{2\pi}{k} e^{-\sigma k^2} e^{i\vc{k}\cdot\vc{\Delta}}
\end{equation}
where $\sigma \equiv \frac{\mu + \nu}{4\mu\nu}$ and $\vc{\Delta} = \vc{R}_{AB} - \vc{R}_{CD}$.  Again we use the scalars $\mu$, $\nu$, and $\sigma$ which correspond to matrices in the more general (non-isotropic) case. Changing to polar coordinates with the $x$-axis along $\vc{\Delta}$,
\begin{eqnarray}
\bra{g_i g_{i'}} \frac{e^2}{\kappa r} \ket{g_j g_{j'}}
&=& \frac{\mathcal{K}e^2}{2\kappa} \frac{\pi}{\mu\nu} \int_0^\pi d\theta\,\int_{-\infty}^\infty dk\,  e^{-\sigma k^2} e^{ik\Delta\cos\theta}\\
&=& \frac{\mathcal{K}e^2}{2\kappa} \frac{\pi}{\mu\nu} \int_0^\pi d\theta\,\int_{-\infty}^\infty dk\,  e^{-\sigma\left(\vc{k}-\frac{i\vc{\Delta}\cos\theta}{2\sigma}\right)^2} e^{-\frac{\Delta^2 \cos^2\theta}{4\sigma}} \label{eqCoulMxElComplSq}\\
&=& \frac{\mathcal{K}e^2}{2\kappa} \frac{\pi}{\mu\nu} \int_0^\pi d\theta\,\left[\int_{-\infty}^\infty dk\,  e^{-\sigma\left(\vc{k}-\frac{i\vc{\Delta}\cos\theta}{2\sigma}\right)^2}\right] e^{-\Delta^2/(8\sigma)}e^{-\frac{\Delta^2 \cos 2\theta}{8\sigma}}  \label{eqCoulMxElTrig}\\
&=& \frac{\mathcal{K}e^2}{2\kappa} \frac{\pi}{\mu\nu} \sqrt{\frac{\pi}{\sigma}} e^{-\Delta^2/(8\sigma)}  \left( \frac{1}{2}\int_0^{2\pi} d\theta'\,e^{-\frac{\Delta^2}{8\sigma}\cos\theta'}\right)  \label{eqCoulMxElPreBessel}\\
&=& \frac{\mathcal{K}e^2}{2\kappa} \frac{\pi^2}{\mu\nu} \sqrt{\frac{\pi}{\sigma}} e^{-\Delta^2/(8\sigma)} I_0\left(\frac{-\Delta^2}{8\sigma}\right)
\end{eqnarray}
On line \ref{eqCoulMxElComplSq} we have completed the square, and to obtain line \ref{eqCoulMxElTrig} the trigonometric identity $\cos^2\theta = (1+\cos 2\theta)/2$ is used.  The quantity in square brackets on line \ref{eqCoulMxElTrig} is equal to $\sqrt{\pi/\sigma}$ (the variable of integration can be shifted).  The parenthesized quantity on line \ref{eqCoulMxElPreBessel} is equal to $\int_0^\pi d\theta' \exp\left[-(\Delta^2/8\sigma)\cos\theta'\right]$ by symmetry, which is equal to $\pi I_0(-\Delta^2/(8\sigma))$ where $I_0$ is the first modified Bessel function.  Combining terms, we arrive at a final analytic formula for Coulomb matrix elements in 2D,
\begin{equation}
\bra{g_i g_{i'}} \frac{e^2}{\kappa r} \ket{g_j g_{j'}} = 
\frac{\mathcal{K}e^2}{2\kappa\sqrt{\sigma}} \frac{\pi^{5/2}}{\mu\nu} e^{-\Delta^2/(8\sigma)}\, I_0\left(\frac{-\Delta^2}{8\sigma}\right) \qquad \mbox{(2D, isotropic)}
\end{equation}

\section{Optimization of Gaussian function parameters\label{appOptimizeGaussians}}
The computational speed of the configuration interaction method described in these notes is dependent on its ability to use relatively small basis sizes (\emph{i.e} numbers of Gaussian elements) to obtain semi-quantitative results.  To achieve a more rapid convergence with respect to the basis size, the placement and exponential factor of the Gaussian basis elements is optimized within a subspace of all possible sets.  This optimization process is of great practical importance, since results must converge before reaching the maximum $n_G$ allowed by our computational resources (currently $50-100$).   In this section, we explain in detail the method used to choose an ``optimal'' set of Gaussian basis functions.
%($\vc{r}_0$ and $\mx{\alpha}$) 

The number of basis elements $n_G$ is always given as a fixed parameter to the optimization procedure.  An initial basis of $n_G$ elements is generated by specifying either (I) the location and size of each dot, along with and the number of elements to place in it, or (II) the location (in a 2D plane) and size of each Gaussian function.  Method (I) requires that the system be comprised of one or more quantum dots, where method (II) can be used for any system.  

Positions of the elements in a Gaussian basis are generated from an underlying two-dimensional mesh of points with two characteristic length scales $a_x$ and $a_y$ (usually the spacing between elements associated with the same quantum dot along the $x$- and $y$-axis, respectively).  Method (I) creates this mesh based on location and size of the dots, and always places a mesh point at the center of each dot. In method (II), the locations of elements are given in terms of $a_x$ and $a_y$, which are given separately.  

The elements in a basis are partitioned into $n_{SS}$ subsets such that the exponential factor, is the same for elements in the same subset.  Initial values for these factors, denoted $\mx{\alpha}_i$ for $i=1\ldots n_{SS}$, are either chosen based on the dot size (method I) or specified directly (method II).  If the Gaussian elements are required to be isotropic, each $\mx{\alpha}$ must be a multiple of the identity.  This restriction results in less freedom for basis optimization but increased computation speed.  In the case of method I, $n_{SS}=2$, and the elements at the center of each dot are allowed to have a different exponential factor than the rest of the elements.  When there are multiple basis elements at a given mesh point the exponential factors of additional elements are found by multiplying the previous element's coefficient by a constant factor $\lambda$.  Method I fixes $\lambda=1.5$ and method II takes a value for $\lambda$ as input.  

Optimization of the basis is performed by minimizing an energy $E$ with respect to $a_x$, $a_y$, $\alpha_i$, and $\lambda$ simultaneously.  The energy minimized is the either (1) the lowest single-electron energy, (2) the lowest many-body energy, or (3) the lowest many-body energy with a given symmetry (\emph{e.g.} $S_z=0$ and total spin $S=1$).  In the results for the exchange energy presented in this work we perform minization via case (3) twice: once to minimize the lowest singlet energy and once to minimize the lowest unpolarized triplet energy.

\end{widetext}

\bibliography{CIpaper}

\end{document}